\documentclass[aps,showpacs,prx,superscriptaddress,preprintnumbers,amsmath,amssymb,twocolumn]{revtex4-2}
\usepackage[utf8]{inputenc}
\usepackage[T1]{fontenc}
\usepackage{graphicx}
\usepackage{dcolumn}
\usepackage{appendix}
\usepackage{amsmath}
\usepackage{bm}
\usepackage[dvipsnames]{xcolor}

\newcommand{\wNN}{\big\lvert w_{v,\overline{\text{NN}}}\big\rangle}
\newcommand{\tpsi}{\big\lvert\Tilde{\psi}(t)\big\rangle}
\newcommand{\tPsi}{\big\lvert\Tilde{\Psi}(t)\big\rangle}
 
\usepackage{subfigure}
\usepackage{bbm}
\usepackage{enumerate}

\usepackage[
  bookmarks=true,
  colorlinks,
  linkcolor=black,
  urlcolor=black,
  citecolor=black,
  plainpages=false,
  pdfpagelabels,
  final,
  breaklinks=true
]{hyperref}

\usepackage{titlesec}

\usepackage{array}

\usepackage{physics}

\begin{document}
\allowdisplaybreaks

\title{Entanglement and non-classical states of light in a strong-laser driven solid-state system}

\author{J.~Rivera-Dean}
\email{javier.rivera@icfo.eu}
\affiliation{ICFO -- Institut de Ciencies Fotoniques, The Barcelona Institute of Science and Technology, 08860 Castelldefels (Barcelona)}

\author{P. Stammer}
\affiliation{ICFO -- Institut de Ciencies Fotoniques, The Barcelona Institute of Science and Technology, 08860 Castelldefels (Barcelona)}

\author{A. S. Maxwell}
\affiliation{Department of Physics and Astronomy, Aarhus University, DK-8000 Aarhus C, Denmark}

\author{Th. Lamprou}
\affiliation{Foundation for Research and Technology-Hellas, Institute of Electronic Structure \& Laser, GR-70013 Heraklion (Crete), Greece}
\affiliation{Department of Physics, University of Crete, P.O. Box 2208, GR-70013 Heraklion (Crete), Greece}

\author{A. F. Ordóñez}
\affiliation{ICFO -- Institut de Ciencies Fotoniques, The Barcelona Institute of Science and Technology, 08860 Castelldefels (Barcelona)}

\author{E. Pisanty}
\affiliation{Department of Physics, King's College London, WC2R 2LS London, United Kingdom}

\author{P. Tzallas}
\affiliation{Foundation for Research and Technology-Hellas, Institute of Electronic Structure \& Laser, GR-70013 Heraklion (Crete), Greece}
\affiliation{ELI-ALPS, ELI-Hu Non-Profit Ltd., Dugonics tér 13, H-6720 Szeged, Hungary}

\author{M. Lewenstein}
\email{maciej.lewenstein@icfo.eu}
\affiliation{ICFO -- Institut de Ciencies Fotoniques, The Barcelona Institute of Science and Technology, 08860 Castelldefels (Barcelona)}
\affiliation{ICREA, Pg. Llu\'{\i}s Companys 23, 08010 Barcelona, Spain}

\author{M. F. Ciappina}
\email{marcelo.ciappina@gtiit.edu.cn}
\affiliation{Physics Program, Guangdong Technion--Israel Institute of Technology, Shantou, Guangdong 515063, China}
\affiliation{Technion -- Israel Institute of Technology, Haifa, 32000, Israel}
\affiliation{Guangdong Provincial Key Laboratory of Materials and Technologies for Energy Conversion, Guangdong Technion – Israel Institute of Technology, Shantou, Guangdong 515063, China}

\date{\today}
\begin{abstract}
    The development of sources delivering non-classical states of light is one of the main needs for applications of optical quantum information science. 
    Here, we demonstrate the generation of non-classical states of light using strong-laser fields driving a solid-state system, by using the process of high-order harmonic generation, where an electron tunnels out of the parent site and, later on, recombines on it emitting high-order harmonic radiation, at the expense of affecting the driving laser field. Since in solid-state systems the recombination of the electron can be delocalized along the material, the final state of the electron determines how the electromagnetic field gets affected because of the laser-matter interaction, leading to the generation of entanglement between the electron and the field. These features can be enhanced by applying conditioning operations, i.e., quantum operations based on the measurement of high-harmonic radiation. We study non-classical features present in the final quantum optical state, and characterize the amount of entanglement between the light and the electrons in the solid. The work sets the foundation for the development of compact solid-state-based non-classical light sources using strong-field physics.
\end{abstract}
\maketitle

Quantum optics stands as one of the most promising natural platforms for the development of practical applications of quantum information science \cite{obrien_photonic_2009,gisin_quantum_2007,gilchrist_schrodinger_2004}, light being the main carrier of information. To pursue such developments, non-classical states of light are needed, i.e. those which are necessarily described by quantum electrodynamical tools \cite{strekalov_nonlinear_2019}. Many of the currently available non-classical light sources rely on nonlinear interactions between light and matter \cite{kwiat_new_1995}. 
Among the most striking examples of nonlinear interactions, is the process of high-order harmonic generation (HHG) \cite{ferray_multiple-harmonic_1988,mcpherson_studies_1987,Brabec-book,villeneuve_attosecond_2018}. Here, a pulsed and strong-laser field interacts with matter leading to the generation of radiation, emitted as a periodic series of ultrashort comb of harmonics of the driving laser that spans dozens, hundreds or even thousands of harmonic orders \cite{popmintchev_bright_2012}. Because of these features, HHG has found numerous applications in attosecond science (cf. \cite{ciappina_attosecond_2017,amini_symphony_2019,krausz_attosecond_2009}), nonlinear XUV optics (cf. \cite{kobayashi_27-fs_1998,midorikawa_xuv_2008,tsatrafyllis_ion_2016,chatziathanasiou_generation_2017,bergues_tabletop_2018,nayak_multiple_2018,senfftleben_highly_2020,orfanos_non-linear_2020}), and high-resolution spectroscopy (cf. \cite{gohle_frequency_2005,cingoz_direct_2012}). In these, a semiclassical theoretical  description of HHG, where light is treated classically while the matter system quantum mechanically \cite{lewenstein_theory_1994,amini_symphony_2019}, is enough to reproduce the experimental results, although leaving unexploited the potential of strong-field physics towards quantum optics \cite{lewenstein_attosecond_2022}. Recently, atomic-HHG processes were studied under a quantum optical framework, proving theoretically and experimentally the generation of non-classical states of light \cite{lewenstein_generation_2021,rivera-dean_strong_2022,stammer_quantum_2022,pizzi_light_2022} spanning from the infrared (IR) up to the extreme ultraviolet (XUV) \cite{stammer_high_2022}. The essential ingredient behind the findings of Refs.~\cite{lewenstein_generation_2021,rivera-dean_strong_2022,stammer_quantum_2022,stammer_high_2022,stammer_theory_2022} was the \emph{conditioning} operation, i.e., restricting measurements of the outgoing radiation to instances in which harmonics were generated. 

In this work, we extend the techniques that have been developed so far for generating non-classical states of light with atomic-HHG processes, to solid-state systems. In atoms, the process behind HHG is resumed in the so-called three-step model \cite{krause_high-order_1992,corkum_plasma_1993,Kulander1993,lewenstein_theory_1994}, in which an electron (i) is extracted from the atom, (ii) accelerates in the continuum driven by the laser field, and (iii) recollides with the parent ion releasing the gained kinetic energy in form of coherent high-frequency radiation. For solids \cite{ciappina_attosecond_2017,kruchinin_colloquium_2018,ghimire_high-harmonic_2019,park_recent_2022}, the process underlying the generation of harmonic radiation becomes more complex \cite{vampa_theoretical_2014,vampa_semiclassical_2015}. Besides the transitions between valence and conduction bands (interband transitions), which find their analogy in gaseous-HHG with the ionization and recombination steps, the electron undergoes excitations within the same band (intraband transitions). This increases the delocalization of the electron, allowing it to recombine, in the emission step later on, with a different site from the one in which it had initially started the dynamics. These additional processes, previously studied under a semiclassical framework \cite{vampa_theoretical_2014,osika_wannier-bloch_2017,PET20}, leave their fingerprints in the final quantum optical state even in the regime of weak delocalization, which we consider here. 
Specifically, the recombination of the electron with a different site from which it was promoted to the conduction band, affects differently the optical state compared to the on-site recombination, leading to electron position-light entanglement. This non-classical effect occurs without any need for conditioning, which might augment the non-classical features, and lead to the generation of coherent state superpositions, as those found in Refs.~\cite{lewenstein_generation_2021,rivera-dean_strong_2022,stammer_high_2022,stammer_quantum_2022}. In Ref.~\cite{gonoskov_nonclassical_2022}, the authors studied, from a quantum optical perspective, the effect of intraband transitions within the conduction band on the final quantum optical state, revealing the presence of non-classical states of light after the measurement of harmonic radiation. Finally, if valence intraband dynamics are very fast, such that recombination may occur anywhere in the solid with a random phase associated, they lead to decoherence of harmonics and hinders the phase matching efficiency, as discussed recently in Ref.~\cite{BJS22}.


\section*{RESULTS}

\begin{figure*}
    \centering
    \includegraphics[width =1\textwidth]{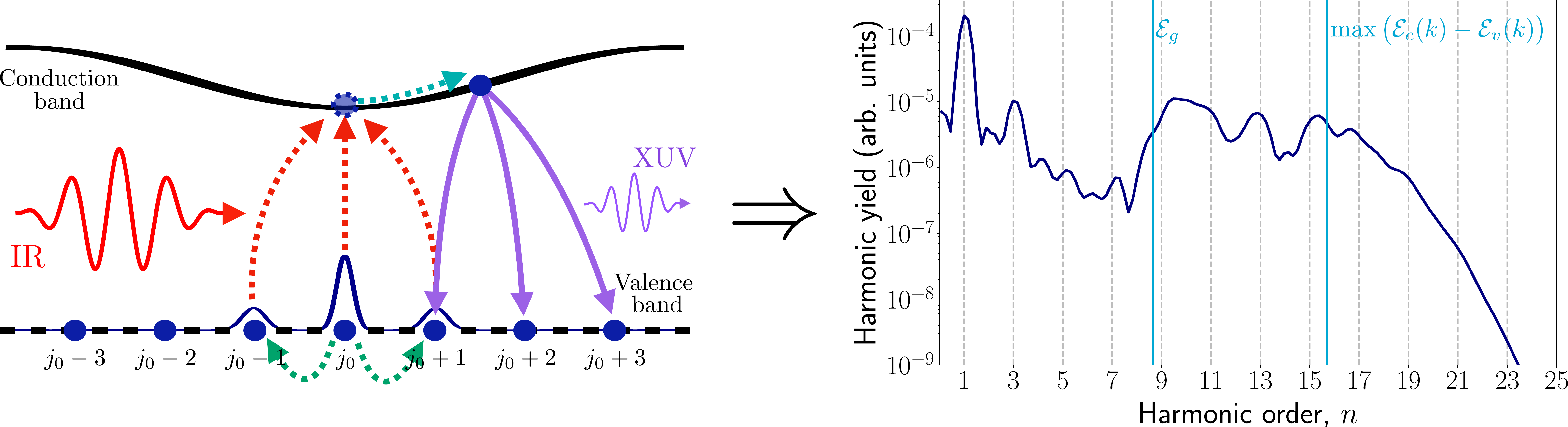}
    \caption{\textbf{Pictorial representation of the high-order harmonic generation process in solid-state systems}. An electron, initially placed at site $j_0$, which can perform intraband transitions to other Wannier sites (green dotted curves), gets excited to the conduction band (red dotted curves). Driven by the input field, the electron accelerates in the conduction band (blue dotted curve) and eventually recombines with a given Wannier site of the valence band (purple solid curves). The energy acquired during its acceleration along the conduction band is emitted in the form of high-harmonic photons, whose spectral distribution is typically characterized by a plateau region ranging from the bandgap energy $\mathcal{E}_g$, up to the maximum energy difference between valence and conduction bands, i.e. $\max(\mathcal{E}_c(k) - \mathcal{E}_v(k))$, as shown in the plot located at the right hand side of the figure.}
    \label{Fig:Scheme}
\end{figure*}

\noindent{\bf Quantum optical framework.} In
typical experimental realizations of intense laser-matter interactions with solid-state systems, a driving laser field of wavelength belonging to the mid-infrared (MIR) regime ($\lambda \sim 3-8$ $\mu$m), hits the target perpendicularly and generates harmonic radiation along the transmission path \cite{chin_extreme_2001,ghimire_observation_2011}. Before the dynamics start, we describe the input electromagnetic field mode with a coherent state $\ket{\alpha_L}$, while all harmonic modes with a vacuum state $\ket{0_{\text{HH}}} = \bigotimes_{q=2} \ket{0_q}$ where $q\in\mathbbm{N}$ denotes the harmonic order. Regarding the solid, we assume it to be in its ground state, corresponding to a completely filled Fermi sea. However, we first solve the time-dependent Schrödinger equation for a single electron, and then phenomenologically treat the many-electron case. Thus, we characterize the initial state of our electron in the Wannier basis as $\ket{w_{v,j_0}}$ (such that $\braket{w_{v,i}}{w_{v,j}} = \delta_{i,j}$), where the subscript $v$ denotes 
the band in which it is located (valence) and $j_0$ its initial site. It can be shown (Methods) that the evolution of the quantum optical state of the system projected onto a Wannier state $\ket{w_{v,j}}$, and neglecting the contributions coming from the conduction band, is given by
\begin{equation}\label{Eq:TDSE:Gen}
    i\hbar \pdv{\ket{\Phi_i(t)}}{t}
        = \sum_{j}\vb{M}_{i,j}(t)\cdot \hat{\vb{E}}(t)
            \ket{\Phi_{j}(t)} \ \forall i \in \mathbbm{Z}.
\end{equation}

Here, $\hbar$ is Planck's constant, $\ket{\Phi_i(t)}$ is the quantum optical state when projected into Wannier state $\ket{w_{v,i}}$, $\vb{M}_{i,j}(t)$ are the matrix elements of the time-dependent dipole moment with respect to $\ket{w_{v,i}}$ and $\ket{w_{v,j}}$, and $\hat{\vb{E}}(t)$ is the time-dependent electric field operator. Considering solids for which the HHG process is localized, such that the electron occasionally ends the dynamics in, at most, its nearest-neighboring sites, we can write the solution to Eq.~\eqref{Eq:TDSE:Gen}, up to first-order perturbation theory, as (Methods)
\begin{equation}\label{Eq:State:init:site}
    \ket{\Phi_{j_0}(t)}
        = e^{i\varphi(t)}\hat{D}_1(\alpha_L) \hat{\mathcal{D}}\big(\boldsymbol{\chi}(t,t_0)\big)
        \bigotimes_{q=1}\ket{0_q},
\end{equation}
when $i = j_0$ in Eq.~\eqref{Eq:TDSE:Gen}, and 
\begin{equation}\label{Eq:State:NN:site}
    \begin{aligned}
    \ket{\Phi_{\text{NN}}(t)}
        &= -\dfrac{i}{\hbar}
            e^{i\varphi(t)}
            \hat{D}_1(\alpha_L) 
            \hat{\mathcal{D}}
                \big(
                    \boldsymbol{\chi}(t,t_0)
                \big)
            \\
            &\quad\times 
            \int^t_{t_0} \dd t' 
                e^{-i\theta(t')}
                \hat{\mathcal{D}}^\dagger\big(\boldsymbol{\chi}(t',t_0)\big)
            \\&\hspace{1.2cm}\times
                \big(
                    \vb{M}_{i,j_0}(t')
                    \cdot
                    \hat{\vb{E}}(t')
                \big)
            \hat{\mathcal{D}}\big(\boldsymbol{\chi}(t',t_0)\big)
            \bigotimes_{q=1}\ket{0_q},
    \end{aligned}
\end{equation}
when $i = j_0\pm1$ in 1D, or generally ``nearest neighbors'', denoted NN. In Eqs.~\eqref{Eq:State:init:site} and \eqref{Eq:State:NN:site}, $\varphi(t)$ and $\theta(t)$ are some phase factors (Methods), $\hat{D}_q(\cdot)$ is the displacement operator acting on mode $q$ \cite{ScullyBook,Gerry__Book_2001}, $\hat{\mathcal{D}}(\boldsymbol{\chi}(t,t_0)) = \prod_{q} \hat{D}_q(\chi_q(t,t_0))$ and $\chi_q(t,t_0)$ is given by
\begin{equation}
    \chi_q(t,t_0)
        = -\dfrac{1}{\hbar}\int^t_{t_0} \dd t' e^{iq\omega_L t'} \vb{g}(\omega_L)\cdot \vb{M}_{i,i}(t'),
\end{equation}
where we have taken into account that all Wannier sites are equivalent and, therefore, $\chi_q(t,t_0)$ is independent of the Wannier site $i$. In these expressions, $\vb{g}(\omega_L)$ arises from the expansion of the electric field operator into the quantized modes.\\

\noindent{\bf Electron position-light entanglement.} The most important difference between the results for solids and atoms regarding the HHG process, is that in solids we encounter the possibility of electron position-light entanglement without the need for \emph{conditioning} operations. We observe that the final Wannier site in which the electron ends up its dynamics, is crucial in determining the final quantum optical state. When valence intraband dynamics are slow on the time scale of the laser pulse duration, both initial and final sites are equal, and results are analogous to those found in atoms \cite{lewenstein_generation_2021,rivera-dean_strong_2022,stammer_quantum_2022}: the different field modes get displaced a quantity $\chi_q(t,t_0)$, corresponding to the Fourier transform of the time-dependent dipole moment averaged over $\ket{w_{v,j_0}}$, that determines the HHG spectrum (Fig.~\ref{Fig:Scheme}). When the electron is found in a different site from $j_0$, the quantum optical state of the system gets affected because of this extra interaction, accounted by $\vb{M}_{i,j_0}(t')\cdot\hat{\vb{E}}(t')$, describing a transition from the initial site $j_0$ to the final site $i$ (in our case $i = j_0\pm 1$) happening at time $t'$. The dynamics at the new site lead to an extra displacement $\chi_q(t,t')$ in each of the modes, indistinguishable from the one we get when recombination happens at site $j_0$, due to the equivalence between different Wannier sites. We note that the electron could potentially hop to other sites, although for the regime considered here, these contributions would be smaller than the ones already introduced, and can be treated with higher-order terms in a perturbation theory-based approach. When valence intraband dynamics are fast enough such that electrons can recombine at any site with a random phase, we get decoherence of HHG radiation and impossibility of efficient phase matching \cite{BJS22}.

For a 1D lattice, the joint state of the system can be written as
\begin{equation}\label{Eq:Lightmatter:entangled}
    \ket{\psi(t)}
        = \ket{w_{v,j_0}}
            \ket{\Phi_{j_0}(t)}
            + \sqrt{2}\ket{w_{v,\text{NN}}}
            \ket{\Phi_{\text{NN}}(t)},
\end{equation}
where we defined $\ket{w_{v,\text{NN}}} = (1/\sqrt{2})(\ket{w_{v,j_0+1}} + \ket{w_{v,j_0-1}})$, and took into account that the quantum optical state $\ket{\Phi_{\text{NN}}(t)}$ is independent of the electron hopping to site $j_0+1$ or $j_0-1$ under the regime of weak delocalization.

If we denote by $f_{\rm c}$ the coordinate number of the lattice (i.e. $f_{\rm c}=2d$ in hypercubic lattices in $d$-dimensions, $f_{\rm c}=3$ for honeycomb lattice in 2D, $f_{\rm c}=6$ for triangular lattice in 2D), in arbitrary dimension we get
\begin{equation}\label{Eq:Lightmatter:entangled1}
    \ket{\psi(t)}
        = \ket{w_{v,j_0}}
            \ket{\Phi_{j_0}(t)}
            + \sqrt{f_{\rm c}}\ket{w_{v,\text{NN}}}
            \ket{\Phi_{\text{NN}}(t)},
\end{equation}
where now $\ket{w_{v,\text{NN}}} = (1/\sqrt{f_{\rm c}})(\sum_{j \in \text{NN}}\ket{w_{v,j_0+j}})$.\\

\begin{figure*}
    \centering
    \includegraphics[width =1\textwidth]{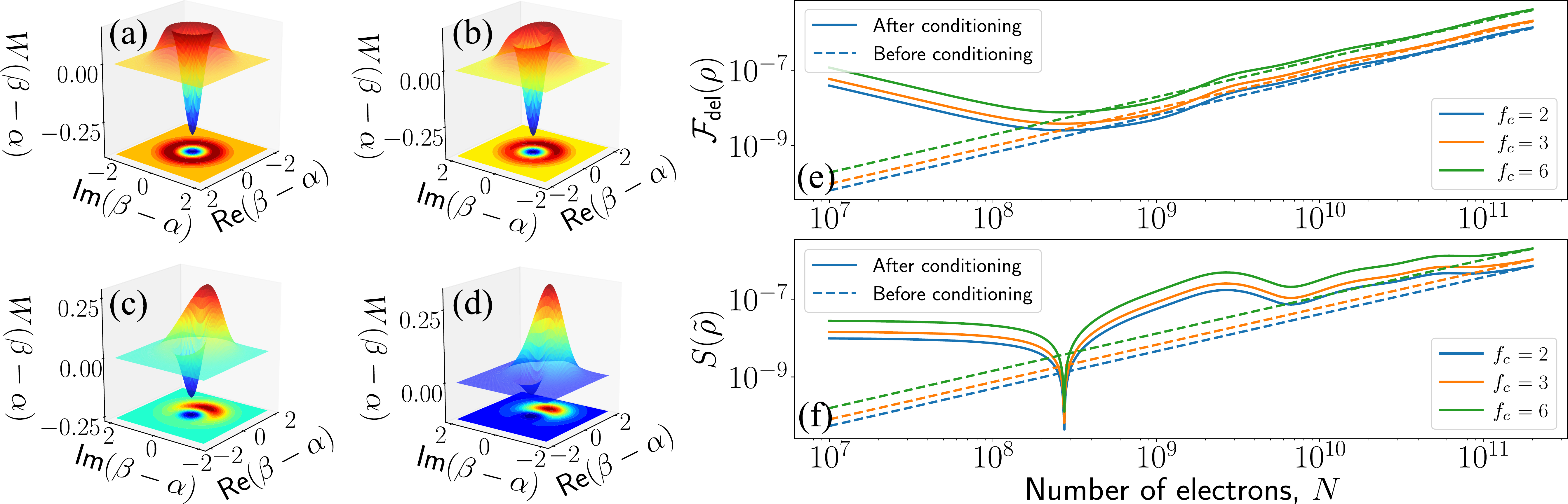}
    \caption{\textbf{Wigner function representation, fidelity and entanglement for the many-electron case}. In (a) to (d), we show the Wigner function distribution of the state after the conditioning for different values of $N$, specifically (a) $N=1 \times 10^7$, (b) $N= 1 \times 10^8$, (c) $N = 5 \times 10^8$ and (d) $N = 1 \times 10^9$. We have used the terminology presented in Ref.~\cite{royer_wigner_1977}, such that $\Re(\beta-\alpha) = x$ and $\Im(\beta-\alpha) = p$ with $x$ and $p$ are the mean value of the quadrature operators $\hat{p} = (\hat{a}-\hat{a}^\dagger)/(i\sqrt{2})$ and $\hat{x} = (\hat{a}+\hat{a}^\dagger)/\sqrt{2}$. In (e) and (f) we show, respectively, the fidelity of the state in Eq.~\eqref{Eq:Lightmatter:entangled:MB} with respect to the state $\wNN$, and the entropy of entanglement $S(\Tilde{\rho})$, with $\Tilde{\rho} = \tr_{\text{field}}(\dyad{\psi(t)})$ for the dashed curves and $\Tilde{\rho} = \tr_{\text{field}}(\dyad{\Psi_{\text{cond}}(t)})$ for the solid curves, when considering different coordinate numbers $f_c$, which correspond to a 1D lattice ($f_c=2$ in blue), a honeycomb lattice in 2D ($f_c=3$ in orange) and a triangular lattice in 2D ($f_c=6$ in green). In both plots, we consider the state before and after the conditioning operation is applied, which are respectively shown with the dashed and solid curves, respectively.}
    \label{Fig:many:electron}
\end{figure*}

\noindent{\bf Phenomenological many-electron theory.} Hitherto, we have only considered the single-electron dynamics. Under the assumption of having localized HHG processes, we can extend the solution given in Eq.~\eqref{Eq:Lightmatter:entangled} to the regime of $N$ independent and phase-matched electrons contributing to the process as (Methods)
\begin{equation}\label{Eq:Lightmatter:entangled:MB}
    \ket{\Psi(t)}
        = \ket{w_{v,\boldsymbol{j_0}}}
            \ket{\Phi_{\boldsymbol{j_0}}(t)}
            + \sqrt{f_c N}\wNN
            \ket{\Phi_{\overline{\text{NN}}}(t)},
\end{equation}
where $|\Phi_{\boldsymbol{j_0}}(t)\rangle$ and $|\Phi_{\overline{\text{NN}}}(t)\rangle$ differ from Eqs.~\eqref{Eq:State:init:site} and \eqref{Eq:State:NN:site} in that the displacement is given as $N\boldsymbol{\chi}(t,t_0)$. In Eq.~\eqref{Eq:Lightmatter:entangled:MB}, $\boldsymbol{j_0}$ is an $N$-dimensional vector describing the initial Wannier site of the electrons, and $\wNN = (1/\sqrt{f_cN})\sum_{\boldsymbol{j}\in\overline{\text{NN}}}\ket{w_{v,\boldsymbol{i}}}$ where $\overline{\text{NN}}:=\{\boldsymbol{j}: \boldsymbol{j_0} \pm (0_1,\cdots,1_i,\cdots,0_N) \ \forall i\}$. The state presented in Eq.~\eqref{Eq:Lightmatter:entangled:MB} has the structure of an entangled state, as recombination in different Wannier sites leads to different quantum optical components. \\

\noindent{\bf Conditioning on HHG.}  Current experimental implementations showed that, quantum operations relying on correlation measurements between the harmonics and part of the fundamental mode \cite{tsatrafyllis_high-order_2017,tsatrafyllis_quantum_2019, lewenstein_generation_2021,rivera-dean_strong_2022,stammer_high_2022,stammer_quantum_2022}, allow for the generation of non-classical states of light \cite{lewenstein_generation_2021,rivera-dean_strong_2022,stammer_high_2022,stammer_theory_2022,stammer_quantum_2022}. Mathematically, this operation can be written by means of the projective operator $\hat{P}_{\text{HHG}} = \mathbbm{1} - \dyad{\alpha}\otimes\dyad{0_{\text{HH}}}$, which we refer to as \emph{conditioning on HHG} operation \cite{stammer_high_2022,stammer_theory_2022}. Applying it onto Eq.~\eqref{Eq:Lightmatter:entangled:MB}, and further projecting with respect to the coherent state $\ket{\boldsymbol{\gamma}} = \bigotimes_{q\geq2}\ket{\gamma_q}$ in which the harmonics are measured, leads
\begin{equation}\label{Eq:Conditioned:MB}
    \ket{\Psi_{\text{cond}}(t)}
        = \braket{\boldsymbol{\gamma}}{\Psi(t)}- \xi(\boldsymbol{\gamma},t)\ket{\alpha_L},
\end{equation}
with $\xi(\boldsymbol{\gamma},t)=\braket{\alpha,0_{\text{HH}}}{\Psi(t)}\braket{\boldsymbol{\gamma}}{0_{\text{HH}}}$. Hereupon, we set $\gamma_q = \chi_q(t,t_0)$.

In general, we have no knowledge about what site has the electron recombined in, and therefore the quantum optical state of the system is given as the mixed state $\hat{\rho}_{\text{field}}(t) = \tr_{\text{elec}}(\dyad{\Psi_{\text{cond}}(t)})$. In the following, we characterize two non-exclusive indications of non-classical behaviors: the presence of Wigner function negativities in $\hat{\rho}_{\text{field}}(t)$, and the presence of entanglement in $\ket{\Psi(t)}$ and $\ket{\Psi_{\text{cond}}(t)}$.

\noindent{\bf Non-classical properties in the many-electron regime.} In the many-electron regime, we consider the case where intraband transitions within the valence band are either negligible, or weak. For HHG processes, it was shown that interband phenomena dominate in the HHG spectrum \cite{vampa_high-harmonic_2015}, which we consider in the computation of $\vb{M}_{i,j}(t)$ (Methods). We use ZnO as a solid system, for which the harmonic emission shows a weaker dependence with the field's ellipticity compared to atoms \cite{ghimire_observation_2011}, suggesting a small influence of delocalized recombination. We excite it with a linearly polarized laser along the $\Gamma-A$ direction, a sin$^2$-shaped envelope, central wavelength $\lambda_L = 3.25 \ \mu\text{m}$, peak intensity $I_0 = 5\times10^{11} \ \text{W/cm}^2$ and $\sim\!40 \ \text{fs}$ of duration.\\

\begin{figure*}
    \centering
    \includegraphics[width =1\textwidth]{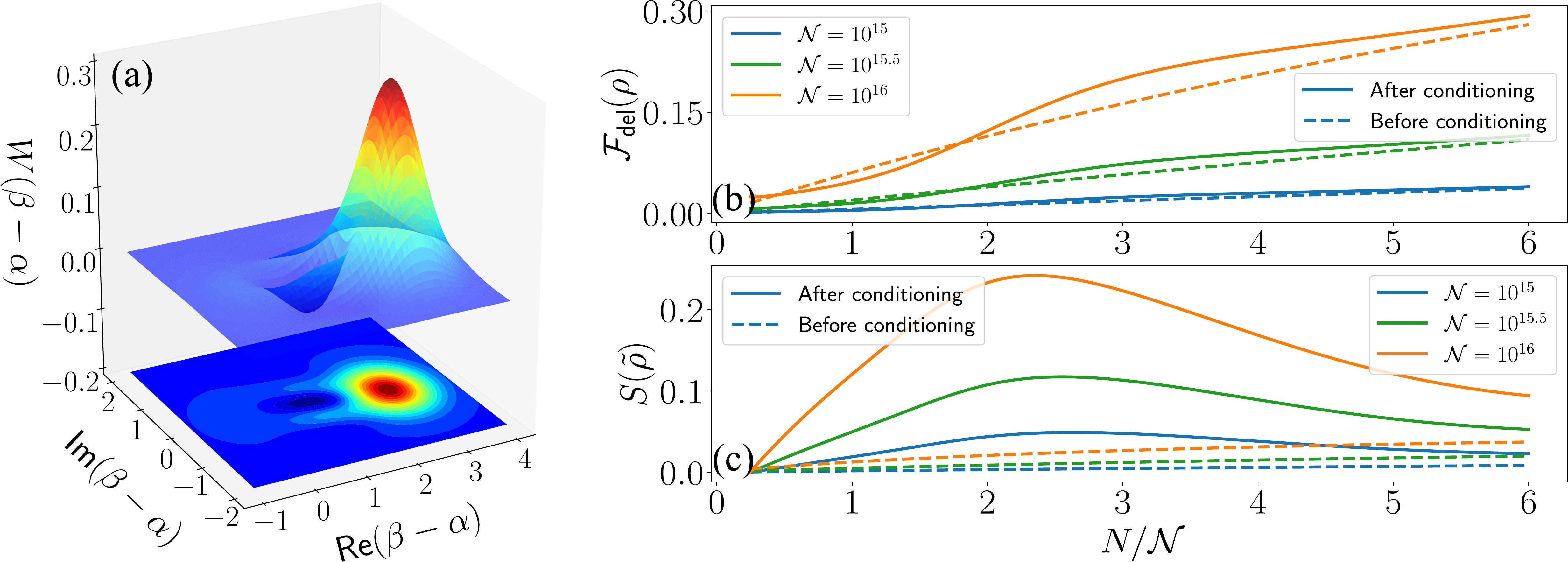}
    \caption{\textbf{Wigner function representation, fidelity and entanglement for the many-electron case when reducing |$\boldsymbol{\chi(t,t_0)}$|}. In (a), we show the Wigner function distribution of the state after the conditioning for $N = 4 \times 10^{16}$. In (b) and (c) we show, respectively, the fidelity of the state in Eq.~\eqref{Eq:Lightmatter:entangled:MB} with respect to the state $\wNN$, and the entropy of entanglement. Here, the $x$ axis represents the number of electrons divided by a quantity $\mathcal{N}$, representing the order of magnitude in the number of electrons. In both plots we consider the state before and after the conditioning operation is applied, which are respectively shown with the dashed and solid curves, respectively.}
    \label{Fig:many:electron:reduced:chi}
\end{figure*}

\noindent{\bf Quantum state of the fundamental after conditioning on HHG.} We study how the Wigner distribution of the quantum optical state in Eq.~\eqref{Eq:Conditioned:MB}, varies when modifying the number of electrons $N$. 
First, we present the results obtained directly from the numerical evaluations, shown in Fig.~\ref{Fig:many:electron}~(a)-(d), where $N\in [1 \times 10^{7},1\times 10^9]$ and $\abs{\boldsymbol{\chi}(t,t_0)} \sim 10^{-3}-10^{-1}$. Similarly to atoms \cite{lewenstein_generation_2021,rivera-dean_strong_2022,stammer_high_2022}, the obtained Wigner distribution shifts from a displaced Fock state (Fig.~\ref{Fig:many:electron}~(a)), to an unbalanced superposition between two close coherent states (Fig.~\ref{Fig:many:electron}~(b) and (c)) until reaching a regime where one of the coherent states in the superposition dominates (Fig.~\ref{Fig:many:electron}~(d)). The Wigner distribution does not present striking differences with respect to those found for atomic systems. This is a consequence of the regime under which we are working, i.e., the dynamics within the valence band are practically negligible. However, note that the properties of $\boldsymbol{\chi}(t,t_0)$ depend on factors such as the nature of the material, the direction along which gets excited, and the characteristics of the field (ellipticity, field intensity and frequency). Thus, if we consider the case where $N = 4\times 10^{16}$, but where $\abs{\boldsymbol{\chi}(t,t_0)}$ is drastically reduced, we get Wigner distributions as the one shown in Fig.~\ref{Fig:many:electron:reduced:chi}~(a). Here, we distinguish two contributions: a big peak around $\text{Re}(\beta-\alpha) \simeq 2.8$, and a very small trough around the origin. While the former corresponds to the shifted part of the state (as in Fig.~\ref{Fig:many:electron}~(d)), the other contribution gets highlighted, as recombinations ending up in the nearest neighbor sites increase with $N$, introducing a small (but non-negligible) vacuum component. Between both, we get Wigner negativities arising from the quantum superposition of these two components.


\noindent{\bf Electron position-light entanglement.} Here, we study the light-matter entanglement as a function of $N$ by computing the entropy of entanglement $S(\Tilde{\rho}):= - \tr_{\text{elec}}(\Tilde{\rho}\log_2(\Tilde{\rho}))$, where $\Tilde{\rho}(t)$ either represents $\tr_{\text{field}}(\dyad{\psi(t)})$ or $\tr_{\text{field}}(\dyad{\Psi_{\text{cond}}(t)})$, depending on whether it is computed before or after the conditioning. We compare the features of this entanglement measure with the ones obtained for the fidelity of $\ket{\psi(t)}$ and $\ket{\Psi_{\text{cond}}(t)}$ with respect to $|w_{v,\overline{\text{NN}}}\rangle$, i.e. $\mathcal{F}_{\text{del}}(\rho) = \text{tr}_{\text{field}}(\langle w_{v,\overline{\text{NN}}}|\hat{\rho}|w_{v,\overline{\text{NN}}}\rangle)$ with $\hat{\rho}=\dyad{\psi(t)}$ or $\hat{\rho} = \dyad{\Psi_{\text{cond}}(t)}$ depending on whether it is computed before or after the conditioning. The results for ZnO are shown in Fig.~\ref{Fig:many:electron}~(b) and (c) for $\mathcal{F}_{\text{del}}(\rho)$ and $S(\Tilde{\rho})$, respectively. We observe that both quantities present very small values, as a consequence of the localized regime under which we are working. However, even in this range, there are some features to highlight. First, for $N \lesssim 2\times10^8$ both the fidelity and the entanglement after the conditioning (solid curves) are bigger than before the conditioning (dashed curves). Here, $\abs{\boldsymbol{\chi}(t,t_0)}$ is a small quantity ($\sim10^{-2}$), allowing the conditioning to highlight contributions of the electron recombining in the nearest-neighbors (shown by the increase in fidelity compared to the situation without conditioning), increasing the value of $S(\Tilde{\rho})$. At $N\simeq 2.5 \times 10^8$, we observe an abrupt drop in $S(\Tilde{\rho})$ after the conditioning, which is absent for $\mathcal{F}_{\text{del}}(\rho)$. Here, the quantum optical contributions arising from the $\lvert w_{\overline{\text{NN}}}\rangle$ term of Eq.~\eqref{Eq:Conditioned:MB}, i.e. $\braket{\boldsymbol{\gamma}}{\Phi_{\overline{\text{NN}}}(t)} - \xi(\boldsymbol{\gamma},t)\ket{\alpha_L}$, cancel each other, leaving the separable on-site contribution, and the observed drop in $S(\Tilde{\rho})$. As $N$ grows, $\xi(\boldsymbol{\gamma},t)$ becomes smaller as $\abs{\boldsymbol{\chi}(t,t_0)}$ gets bigger, since the overlap with the initial state is very small. In this regime, where conditioning does nothing, $\mathcal{F}_{\text{del}}(\rho)$ and $S(\Tilde{\rho})$ for the state before and after the conditioning coincide. Note that these two quantities become bigger as the coordinate number $f_c$ increases. If the same analysis is done for the case where $N$ is increased to the order of $10^{16}$ while $\abs{\boldsymbol{\chi}(t,t_0)}$ is drastically reduced, $S(\Tilde{\rho})$ and $\mathcal{F}_{\text{del}}(\rho)$ get enhanced (Fig.~\ref{Fig:many:electron:reduced:chi}~(b) and (c)). When $\abs{\boldsymbol{\chi}(t,t_0)} \sim 10^{-1}$, we observe that $S(\Tilde{\rho})\simeq 0.2$ for $N \simeq 2.2\times 10^{16}$, while $\mathcal{F}_{\text{del}}(\rho)\simeq 0.15$. For other values of $N$, we observe the same behaviour as presented in Figs.~\ref{Fig:many:electron}~(e) and (f): for $N < 2.2\times 10^{16}$, $S(\Tilde{\rho})$ drops to zero, which identifies with the huge dip observed in Fig.~\ref{Fig:many:electron}~(f), while below this value a small enhancement is observed, although not shown in this plot; for $N > 2.2\times 10^{16}$, both $S(\Tilde{\rho})$ and $\mathcal{F}_{\text{del}}(\rho)$ before and after the conditioning converge.

\begin{figure*}
    \centering
    \includegraphics[width =1\textwidth]{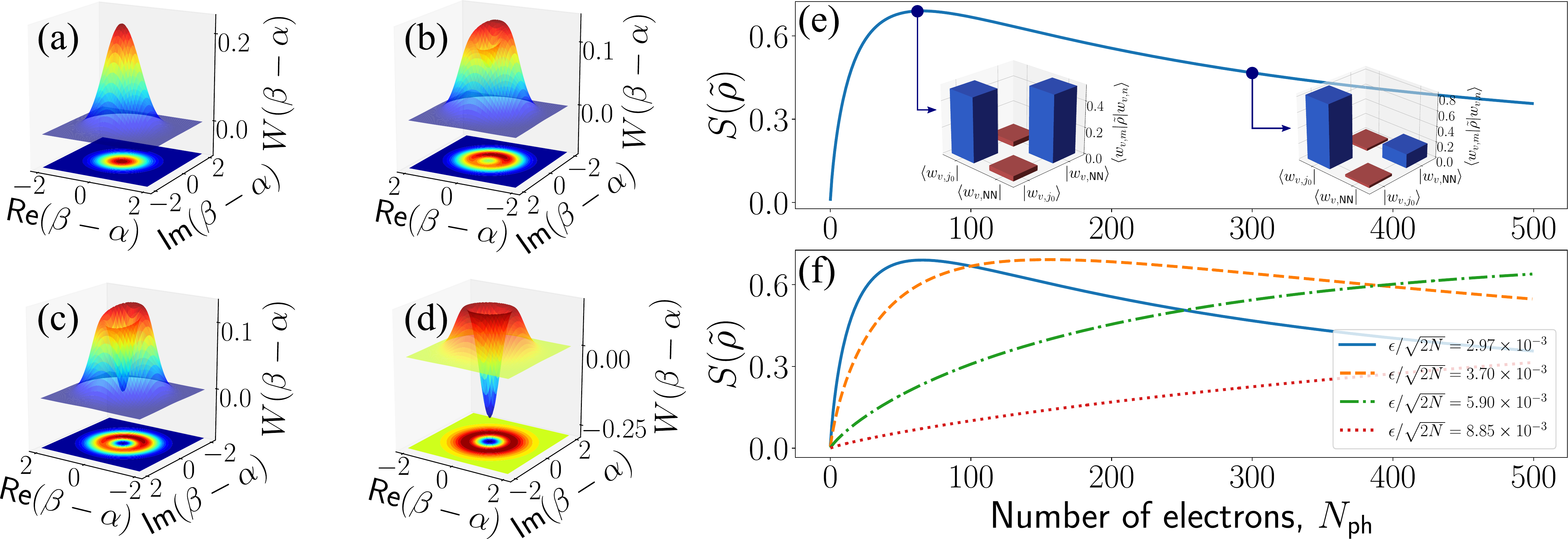}
    \caption{\textbf{Wigner function representation and many-electron entanglement for the few-electron case}. In (a) to (d), we show the Wigner function representation of the state after the conditioning for different values of $N$, specifically (a) $N=10$, (b) $N= 40$, (c) $N = 60$ and (d) $N = 300$, while in all cases we have set $\epsilon/\sqrt{2 N} = 2.97\times 10^{-3}$, where we have considered a 1D lattice such that $f_c = 1$. In (e) and (f) we show the entropy of entanglement as a function of the number of electrons. While in (e) we have set $\epsilon/\sqrt{2 N} = 2.97\times 10^{-3}$, in (f) we have considered different values of $\epsilon$ satisfying $\epsilon\sqrt{2N} \leq 0.3$. The histograms shown in the inset of (e) present the norm of the matrix elements $\lvert\mel{w_{v,m}}{\Tilde{\rho}}{w_{v,n}}\rvert$, with $\Tilde{\rho}$ the state of the system once the partial trace with respect to the quantum optical degrees of freedom has been done.}
    \label{Fig:few:electron}
\end{figure*}


\noindent{\bf Non-classical properties in the few-electron regime.} Here, we consider the few-electron regime and, in counterpart, we increase the intraband dynamics within the valence band by defining a perturbation parameter $\epsilon$ (Methods) satisfying $\epsilon\sqrt{f_cN}<0.3$. We note that an increase of the valence band dynamics corresponds to an enhancement of the hopping potential between sites $I_v$, which is related with the curvature of the valence band (Methods). Therefore, the localized regime corresponds to solids for which the valence band is essentially flat, i.e. $I_v\lesssim 10^{-2}$, when excited with pulses of $\sim 40 \ \text{fs}$ of duration.

Fig.~\ref{Fig:few:electron}~(a)-(d) shows the Wigner function after the conditioning, respectively for $N=10,40,60,300$, for which $\abs{\boldsymbol{\chi}(t,t_0)}\sim 10^{-8}-10^{-6}$, while setting $\epsilon/\sqrt{2N} = 2.97\times 10^{-3}$. Unlike in the many-electron regime and atomic systems, for small values of $N$ we get a Gaussian distribution centered at $\beta = \alpha$ (Fig.~\ref{Fig:few:electron}~(a)). As mentioned, when decreasing the value of $N$, the conditioning operation highlights the contributions coming from the delocalized part of the state. However, here we also increase the contribution of delocalized processes, so their contribution is still significant after conditioning within the considered range of $N$. For small values of $N$ the most important contribution comes from the $\lvert w_{\overline{\text{NN}}}\rangle$ term in Eq.~\eqref{Eq:Conditioned:MB}, and the Wigner function corresponds to a Gaussian. By increasing $N$, this situation reverses, and the Wigner function presents a small minimum at the center, which becomes deeper for bigger $N$ (Figs.~\ref{Fig:few:electron}~(b) and ~(c)). When $N$ becomes sufficiently large, the $\lvert w_{\boldsymbol{j_0}}\rangle$ contribution dominates and we recover the Wigner function of a displaced Fock state, with a huge negative minimum at the center (Fig.~\ref{Fig:few:electron}~(d)).

This transition where the delocalized contribution becomes less important, has striking consequences on $S(\Tilde{\rho})$, shown in Figs.~\ref{Fig:few:electron}~(e) and (f) for different values of $\epsilon$. We observe that for $N \approx 60$ and $\epsilon/\sqrt{2N} = 2.97\times 10^{-3}$, $S(\Tilde{\rho})\simeq 0.7$ (Fig.~\ref{Fig:few:electron}~(e)), when the localized and delocalized terms contribute equally to the quantum state of the system (inset plot at the left of Fig.~\ref{Fig:few:electron}~(e)). For increasing $N$, the amount of entanglement decays as the localized contribution to the state becomes more important (inset at the right of Fig.~\ref{Fig:few:electron}~(e)). For increasing values of $\epsilon$, i.e. as more important delocalization processes become, the maximum of the entanglement entropy shifts to bigger $N$ (Fig.~\ref{Fig:few:electron}~(f)). Thus, the stronger the electrons interact in the material, the higher the correlations established with the field become.

\section*{DISCUSSION}
This work extends the formalism and techniques that allow generating non-classical states of light in atomic systems that undergo HHG processes, to solid-state targets. In these, and within the regime considered along the work, the electron not only recombines with the site from which it got promoted to the conduction band, but also with its nearest neighbors. This leads to the generation of entangled states between the electron's final position and the light. In the case where the recombination is highly localized, the amount of generated entanglement is almost negligible, although the performance of conditioning operations, which allows generating non-classical states of light \cite{lewenstein_generation_2021,rivera-dean_strong_2022,stammer_high_2022,stammer_theory_2022,stammer_quantum_2022}, enhances these characteristics. We also considered two ways of increasing these features: one in which the number of electrons participating in the process is huge, but the generated light shift (depending on the material, direction along which it has been excited and field parameters) is drastically reduced; the other, in which the number of electrons gets reduced, but in counterpart the interaction between the electrons in the valence band increases (in the regime where the valence band is almost flat).

Generating quantum correlations between different parties, is a key aspect for applications of quantum information science such as quantum communication \cite{gisin_quantum_2007,brunner_bell_2014,nadlinger_experimental_2022,zhang_device-independent_2022} and quantum computation \cite{cirac_quantum_1995,knill_scheme_2001,clarke_superconducting_2008}. Strong-field physics could extend these to unprecedented time and energy scales \cite{stammer_high_2022}. We have seen that the use of solid-state systems could allow for establishing correlations between electrons and light. We have studied this in the case where the interaction between electrons is weak, although non-vanishing. A potential extension to this work could consider the case where the electrons strongly interact with their surroundings, leading to a highly delocalized recombination which might greatly affect the final quantum optical state. Furthermore, in a different direction, Ref.~\cite{pizzi_light_2022} recently showed that introducing quantum correlations in the initial state of an atomic ensemble, could lead to non-classical states of light after HHG processes, and to correlations between the harmonic modes. Thus, as a more general scope, these works aim at connecting quantum optics with attosecond science with the purpose of realizing sources for investigations on quantum information processing \cite{lewenstein_attosecond_2022}, working at atomic native time and energy scales.


\bibliography{References.bib}{}

\section*{METHODS}
\subsection*{Solving the time-dependent Schrödinger equation}
The Hamiltonian characterizing the dynamics of the total system is given, under the dipole approximation and within the length gauge \cite{stammer_quantum_2022}, by
\begin{equation}\label{Eq:total:Hamiltonian}
    \hat{H}
        = \hat{H}_{\text{cr}} 
        + e\hat{\vb{R}}\cdot \hat{\vb{E}}(t)
        + H_{\text{f}},
\end{equation}
where $e$ is the electron's charge, $\hat{H}_{\text{cr}}$ is the crystal Hamiltonian, $\vb{\hat{R}}$ is the electron's position operator, $\hat{\vb{E}} = -i \sum_{q} \vb{g}(\omega_q)(\hat{a}^\dagger_q - \hat{a}_q)$ is the field operator with $\hat{a}_q$ ($\hat{a}^\dagger_q$) the annihilation (creation) operator acting over the $q$th mode ($q \in \mathbbm{N}$), $\vb{g}(\omega_q)$ is a factor arising from the expansion of the electric field operator into the quantized modes, and $\hat{H}_{\text{f}} = \sum_q \hbar \omega_q \hat{a}^\dagger_q \hat{a}_q$ is the free-field Hamiltonian.

For the numerical analysis, we consider a two-band model under the tight-binding approximation, and describe the energy dispersion relations for the valence and conduction bands, respectively, as \cite{HaugBook}
\begin{align}
    &\mathcal{E}_v(k) 
        = -2I_v\cos(k\mathsf{a}),\\
    &\mathcal{E}_c(k)
        = \mathcal{E}_c' - 2I_c\cos(k\mathsf{a}),
\end{align}
where $k$ is the crystal momentum, $\mathsf{a}$ is the lattice constant, $I_v < 0$ and $I_c > 0$ are the hopping parameters in the valence and conduction bands, respectively, and $\mathcal{E}_c' = \mathcal{E}_g + 2I_c - 2I_v$, with $\mathcal{E}_g$ the band-gap energy.

In order to solve the Schrödinger equation in \eqref{Eq:total:Hamiltonian}, we first perform some unitary transformations that simplify the analysis. First, we move to the interaction picture with respect to the free-field term, i.e. $\ket{\psi(t)} = e^{iH_\text{f}t}\ket{\psi'(t)}$ which leads to a time dependence in the electric field operator ($\hat{a}_q^\dagger \to \hat{a}_q^\dagger e^{i\omega_q t}$). Secondly, we work in a displaced frame $\ket{\psi'(t)} = D(\alpha_L) \ket{\psi''(t)}$ such that the electric field operator splits into a classical part $\vb{E}_{\text{cl}}(t) = \mel{\alpha_L,0_{\text{HH}}}{\hat{\vb{E}}(t)}{\alpha_L,0_{\text{HH}}}$, with $\ket{0_{\text{HH}}} = \bigotimes_{q=2}\ket{0_q}$, and another term $\hat{\vb{E}}(t)$ describing the quantum fluctuations. Thus, at this level the Hamiltonian reads
\begin{equation}
    \hat{H}''(t) = \hat{H}_{\text{cr}}
            + e\hat{\vb{R}}\cdot \vb{E}_{\text{cl}}(t)
            + e \hat{\vb{R}}\cdot\hat{\vb{E}}(t),
\end{equation}
and the initial state of the system is $\ket{\psi''(t_0)} = \ket{w_{v,j_0}}\bigotimes_{q=1} \ket{0_q}$. Finally, we work in the interaction picture with respect to the semiclassical Hamiltonian $\hat{H}_{\text{sc}}(t) = \hat{H}_{\text{cr}} + e \hat{\vb{R}}\cdot \vb{E}_{\text{cl}}(t)$, i.e. $\ket{\Psi''(t)} = U_{\text{sc}}(t)\tpsi$ with $U_{\text{sc}}(t) = \mathcal{T}\exp[-\tfrac{i}{\hbar}\int^t_{t_0}\dd t' \hat{H}_{\text{sc}}(t')]$, such that the Schrödinger equation reads
\begin{equation}
    i\hbar \dv{\tpsi}{t}
        = e \hat{\vb{R}}(t)\cdot\hat{\vb{E}}(t) \tpsi,
\end{equation}
where $e\hat{\vb{R}}(t) = eU^\dagger_{\text{sc}}(t)\hat{\vb{R}}U_{\text{sc}}(t)$ is the time-dependent dipole operator.

At this point, we introduce the identity under a Wannier-Bloch mixed representation \cite{osika_wannier-bloch_2017}, where we consider  Wannier states $\ket{w_{v,j}}$ for the valence band and Bloch states for the conduction band
\begin{equation}
    \mathbbm{1} 
        = \sum_{j} \dyad{w_{v,j}}
            + \int \dd k \dyad{\phi_{c,k}},
\end{equation}
and in a similar approach as those in Refs.~\cite{lewenstein_generation_2021,rivera-dean_strong_2022,stammer_high_2022,stammer_theory_2022}, we neglect the contributions of the conduction band terms to the dynamics. Although these studies are based in atomic systems, we expect this approximation to hold in solid state systems as well, as the frequency of the applied pulse is much smaller than the bandgap energy of the solid \cite{osika_wannier-bloch_2017}, and therefore the conduction band gets hardly populated at the end of the pulse. Moreover, in solids we have decoherence effects which make the excited electron to eventually return to the valence band, decreasing even further the probability of finding an excited electron in the valence band \cite{BJS22,wang_quantum_2021}. However, we remark that decoherence effects are not taken into account in our equations. Thus, under this approximation, we have
\begin{equation}
    i\hbar \dv{\tpsi}{t}
        \approx e\sum_j\hat{\vb{R}}(t)\cdot\hat{\vb{E}}(t) \big\langle w_{v,j}\tpsi\ket{w_{v,j}}.
\end{equation}

By projecting this equation onto the set of Wannier states $\{\ket{w_j}: j\in \mathbbm{Z}\}$, we get the following system of differential equations
\begin{equation}\label{Eq:solids:diff:eqs}
    i \hbar \dv{\ket{\Phi_i(t)}}{t}
        = \sum_j \vb{M}_{i,j}(t)\cdot\hat{\vb{E}}(t) \ket{\Phi_{j}(t)}
        \ \forall i \in \mathbbm{Z},
\end{equation}
where we have defined $\ket{\Phi_i(t)} = \big\langle w_{v,i}\tpsi$ and $\vb{M}_{i,j}(t) = e\mel{w_{v,i}}{\hat{\vb{R}}(t)}{w_{v,j}}$. Thus, we encapsulate the electron dynamics in $\vb{M}(t)$. In systems where the electron does not interact with its surroundings, and for which the initial and final state after HHG is the same, $\vb{M}(t)$ would be a diagonal matrix. However, in solid-state systems, this is in general not the case, and we thus get non-vanishing values for the off-diagonal elements in $\vb{M}(t)$. For instance, solid ZnO presents a much weaker dependence of the high-harmonic yield with the driving field ellipticity compared to what is found in atomic systems \cite{ghimire_observation_2011}, a signature that suggests the presence of delocalized processes in solid-HHG, i.e.~the final state of the electron does not need to be the same as the starting one. 

Here, we investigate the quantum optical features of the final HHG state under the assumption that the HHG process is weakly delocalized, i.e., $\abs{\vb{M}_{i,j}(t)} \ll \abs{\vb{M}_{i,i}(t)}$ with $i\neq j$. Specifically, under this regime we note that the matrix elements $\abs{\vb{M}_{i,j}(t)}$ are smaller the bigger the distance $\abs{i-j}$ is. Here, we only consider nearest-neighbor contributions, i.e. $j = i\pm 1$, and treat them as a perturbation parameter $\epsilon$. Thus, we consider a perturbation theory expansion of $\ket{\Phi_{i}(t)}$ up to first order in $\epsilon$
\begin{equation}
    \ket{\Phi_i(t)}
        \approx \ket{\Phi^{(0)}_i(t)}
            + \ket{\Phi_i^{(1)}(t)},
\end{equation}
such that the zeroth order term satisfies
\begin{equation}
    i\hbar \dv{\ket{\Phi^{(0)}_i(t)}}{t} 
        = \vb{M}_{i,i}(t)\cdot \hat{\vb{E}}(t) \ket{\Phi^{(0)}_i(t)},
\end{equation}
and whose solution can be written as \cite{lewenstein_generation_2021,rivera-dean_strong_2022,stammer_quantum_2022}
\begin{equation}
    \begin{aligned}
    \ket{\Phi^{(0)}_i(t)}
        &= \hat{\mathcal{D}}\big(\boldsymbol{\chi}_{i,i}(t,t_0)\big)
            \ket{\Phi_i^{(0)}(t_0)}
        \\
        &\equiv \prod_q e^{i\varphi^{(q)}_{i,i}(t,t_0)}
            \hat{D}_q
                \big(
                    \chi^{(q)}_{i,i}(t,t_0)
                \big)
                \ket{\Phi_i^{(0)}(t_0)}
    \end{aligned}
\end{equation}
where $\hat{D}_q(\alpha) = \exp[\alpha\hat{a}_q^\dagger-\alpha^*\hat{a}_q]$ is the displacement operator acting on mode $q$ \cite{ScullyBook,Gerry__Book_2001}, $\varphi^{(q)}_{i,i}(t)$ is a prefactor arising from applying the BCH relation (see Refs.~\cite{lewenstein_generation_2021,rivera-dean_strong_2022,stammer_quantum_2022} for details), and $\chi^{(q)}_{i,i}(t,t_0)$ is a coherent state amplitude given by
\begin{equation}
    \chi^{(q)}_{i,i}(t,t_0)
        = -\dfrac{1}{\hbar}
            \int^t_{t_0}
                \dd\tau e^{i\omega_q \tau}
                \vb{g}(\omega_q)\cdot\vb{M}_{i,i}(\tau).
\end{equation}

Note that, unlike the main text, here we have explicitly added the dependence of $\chi^{(q)}_{i,i}(t,t_0)$ with the Wannier site along which the dynamics take place, in this case, the $i$th site. However, we note that dynamics in all the sites are equivalent, and therefore they lead to the same high-harmonic spectra, allowing us to drop the $i$ index in $\chi^{(q)}_{i,i}(t,t_0)$. However, for the sake of completeness, we keep this notation in the following.

On the other hand, for the first-order term we get the following differential equation
\begin{equation}
    \begin{aligned}
    i\hbar \dv{\ket{\Phi^{(1)}_i(t)}}{t} 
        &= \vb{M}_{i,i}(t)\cdot \hat{\vb{E}}(t) \ket{\Phi^{(1)}_i(t)}
        \\&\quad
            + \sum_{j\in\{i\pm1\}} \vb{M}_{i,j}(t)\cdot \hat{\vb{E}}(t)
                \ket{\Phi^{(0)}_j(t)},
    \end{aligned}
\end{equation}
and whose solution is, in general, given by
\begin{equation}
    \begin{aligned}
    \ket{\Phi^{(1)}_i(t)}
        &= \hat{\mathcal{D}}
                    \big(
                        \boldsymbol{\chi}_{i,i}(t,t_0)
                    \big)
            \ket{\Phi_i^{(1)}(t_0)}
            \\& \quad
            - \dfrac{i}{\hbar}
                \sum_{j\in \{i\pm 1\}}
                    \int^t_{t_0}
                        \dd t' \hat{\mathcal{D}}
                                \big(
                                    \boldsymbol{\chi}_{i,i}(t,t')
                                \big)
                            \vb{M}_{i,j}(t')\cdot\hat{\vb{E}}(t')
                                \\&\hspace{2.5cm}\times
                               \hat{\mathcal{D}}
                                    \big(
                                        \boldsymbol{\chi}_{j,j}(t',t_0)
                                    \big)
                                    \ket{\Phi^{(0)}_j(t_0)}.
    \end{aligned}
\end{equation}

Thus, and taking into account the initial conditions, we find two kinds of contributions depending on the final site $j$ where the electron ends up in:
\begin{itemize}
    \item If the electron ends up the dynamics in the initial Wannier site $j_0$, then the final state reads
    \begin{equation}\label{Eq:On:Site:Contrib}
        \ket{\Phi_{j_0}(t)}
            = \hat{\mathcal{D}}
                \big(
                    \boldsymbol{\chi}_{i,i}(t,t_0)
                \big)\bigotimes_q \ket{0_q};
    \end{equation}
    \item If the electron ends up the dynamics in a different Wannier site $j$ from which it started, then the final state becomes
    \begin{equation}\label{Eq:Diff:Site:Contrib}
        \begin{aligned}
        \ket{\Phi_{j}(t)}
            = -\dfrac{i}{\hbar}
                    \int^t_{t_0}
                        \dd t' &\hat{\mathcal{D}}
                                \big(
                                    \boldsymbol{\chi}_{j,j}(t,t')
                                \big)
                            \vb{M}_{j,j_0}(t')\cdot\hat{\vb{E}}(t')
                                \\&\times
                               \hat{\mathcal{D}}
                                    \big(
                                        \boldsymbol{\chi}_{j_0,j_0}(t',t_0)
                                    \big)
                                    \bigotimes_q \ket{0_q}.
        \end{aligned}
    \end{equation}
\end{itemize}

Note that, by taking into account that $\chi^{(q)}_{i,i}(t)=\chi^{(q)}_{j,j}(t) = \chi_q(t)$, the latter given in the main text, we can rewrite Eq.~\eqref{Eq:Diff:Site:Contrib} as
\begin{equation}
    \begin{aligned}
    \ket{\Phi_{j}(t)}
        &= -\dfrac{i}{\hbar}
            e^{i\varphi(t)}
            \hat{\mathcal{D}}
                \big(
                    \boldsymbol{\chi}(t,t_0)
                \big)
            \\
            &\quad\times 
            \int^t_{t_0} \dd t' 
                e^{-i\theta(t')}
                \hat{\mathcal{D}}^\dagger\big(\boldsymbol{\chi}(t',t_0)\big)
            \\&\hspace{1.2cm}\times
                \big(
                    \vb{M}_{j,j_0}(t')
                    \cdot
                    \hat{\vb{E}}(t')
                \big)
            \hat{\mathcal{D}}\big(\boldsymbol{\chi}(t',t_0)\big)
            \bigotimes_{q=1}\ket{0_q},
    \end{aligned}
\end{equation}
where we have defined $\theta(t') = \sum_q\Im{\chi_q(t,t_0)\chi_q^*(t',t_0)}$. Furthermore, we find that for the regime of weak delocalization, under which only the nearest-neighbor sites get populated, the integrals appearing in the expression above are equal up to the fifth significant decimal (in the worst-case scenario for which we enhance the intraband transitions) when $j=\pm 1$.

\subsection*{Semiclassical approach under a Wannier-Bloch picture}
Along the main text, we characterize the valence band states via a Wannier representation. As a consequence, the final quantum optical state gets affected by how the dynamics of the electron behave in this picture, via the matrix elements of $\vb{M}(t)$ given by
\begin{equation}\label{Eq:mel:sc}
    \vb{M}_{i,j}(t)
        = \mel{w_{v,i}}{U_{\text{sc}}^\dagger(t,t_0)
        \hat{\vb{R}}U_{\text{sc}}(t,t_0)}{w_{v,j}}.
\end{equation}

Here, we are going to present the main expressions obtained in Ref.~\cite{osika_wannier-bloch_2017} (further investigated in Ref. \cite{PET20}), where a semiclassical study of solid-HHG is performed under a Wannier-Bloch approach, that allows us to later compute the matrix elements in Eq.~\eqref{Eq:mel:sc}. Assuming a one-dimensional system, and that the electron begins the dynamics in site $j_0$, according to Ref.~\cite{osika_wannier-bloch_2017} at time $t$ we get for the time-evolved quantum state
\begin{equation}\label{Eq:ansatz:sc}
    \ket{\psi(t)}
        = \sum_j a_{j,j_0}(t)
            \ket{w_{v,j}}
        + \int_{\text{BZ}} \dd k \ a_{c}(k,t)
            \ket{\phi_{c,k}},
\end{equation}
where $a_{j,j_0}(t)$ is the probability amplitude of an electron being in site $j$ at time $t$ given that was initially at site $j_0$, and $a_{c}(k,t)$ is the probability amplitude of being in the valence band with crystal momentum $k$. The expression for these probability amplitudes, under the tight-binding approximation and for nearest-neighbor interactions, become
\begin{equation}
    \begin{aligned}
    a_{j,j_0}(t)
        = \sum_q &e^{iq(j-j_0)}
            \\&\times
                \exp[2\dfrac{i}{\hbar}I_v \cos(q+\dfrac{e \mathrm{a}}{\hbar c}A(t) -\dfrac{e \mathrm{a}}{\hbar c}A(\tau))],
    \end{aligned}
\end{equation}
and
\begin{equation}\label{Eq:conduction:sc}
    \begin{aligned}
    a_{c}(p,t)
        = -\dfrac{i}{\hbar}
            \sum_j \int^{t}_{t_0} &\dd t'
                \int_{\widetilde{\text{BZ}}} \dd p
                E_{\text{cl}}(t')a_{j,j_0}(t')
                \\&\times
                d^*_{jc}
                    \Big(
                        p-\dfrac{e}{c}A(t')
                    \Big)
                \\&\times
                \exp[-\dfrac{i}{\hbar} \int^t_{t'}\dd \tau
                        \mathcal{E}_c
                            \big(
                                p-\tfrac{e}{c}A(\tau)
                            \big)],
    \end{aligned}
\end{equation}
where in this last expression $p = k + \tfrac{e}{c}A(t)$ is the canonical momentum and $\widetilde{\text{BZ}} = \text{BZ}+\tfrac{e}{c}A(t)$ is the shifted Brillouin zone. Furthermore, we have defined $d_{j,c}(k) = e\mel{w_{v,j}}{\hat{X}}{\phi_{c,k}}$.


By combining Eqs.~\eqref{Eq:mel:sc} and \eqref{Eq:ansatz:sc}, we find that the matrix elements of the transition matrix $M(t)$ are given by
\begin{equation}\label{Eq:Matrix:elements}
    \begin{aligned}
    M_{i,j}(t)
        &= \sum_{m,n} a^*_{m,i}(t) a_{n,j}(t)
                    \mel{w_{v,m}}{\hat{X}}{w_{v,n}}
        \\&\quad
         + e\sum_{m} \int_{\text{BZ}} \dd k 
            a^*_{m,i}(t)a_{c}(k,t)
            \mel{w_{v,m}}{\hat{X}}{\phi_{c,k}}
        \\&\quad
         + e\sum_{n} \int_{\text{BZ}} \dd k 
            a^*_{c}(k,t) a_{n,j}(t)
            \mel{\phi_{c,k}}{\hat{X}}{w_{v,n}}
        \\&\quad
        + e\int_{\text{BZ}} \dd k 
            \int_{\text{BZ}} \dd k' 
                a^*_{c}(k',t) a_{c}(k,t)
            \mel{\phi_{c,k'}}{\hat{X}}{\phi_{c,k}},
    \end{aligned}
\end{equation}
where we note that $M_{i,j}(t) = M^*_{j,i}(t)$. Each of the contributions to the matrix elements describe different processes. The first term describes an intraband transition between two Wannier sites in the valence band; the second and third terms, describe an interband transition between the valence band and a given site of the conduction band; finally, the fourth term describes an intraband transition within the conduction band. Additionally, we note that at $t = t_0$ the matrix is diagonal, i.e. $M_{i,j}(t) \propto \delta_{i,j}$. For $t>t_0$ we may get additional off-diagonal contributions depending on the characteristics of the considered solid. 

We note that, for $i=j$ and in the limit of weak-delocalization, the first term in \eqref{Eq:Matrix:elements} remains almost constant as for 1D systems one has $\mel{w_{v,i}}{\hat{X}}{w_{v,j}} = x_j \delta_{i,j}$ (see Supplementary Material), and therefore can be written as $\sum_m x_m \abs{a_{m,j}(t)}^2$. On the other hand, for $i\neq j$ we get instead $\sum_m x_m a_{m,j}(t)a^*_{m,i}(t)$, such that at $t = t_0$ the $i=j$ contribution survives while the $i\neq j$ does not. For the many-electron regime, we neglect this contribution at all times, arguing that the delocalization effects due to intraband dynamics in the valence band are very small. Instead, in the few-electron regime we considered this quantity to be small (in agreement with the perturbation approach that we have developed) but non-negligible. Thus, we labeled this term as $\epsilon$ (see Fig.~\ref{Fig:few:electron}), and increased it in the numerical analysis, in an equivalent way as if we were considering solids for which the valence band is essentially flat. Specifically, this is valid for valence band hopping parameter $I_v < 10^{-2}$ when considering laser pulses of $\sim 40$ fs of duration, or lower. Of course, if laser pulses of greater duration are employed, then the electron could potentially end up in other Wannier sites different from the nearest neighbor ones, but we do not contemplate this case in our calculations.

\subsection{Solving the Schrödinger equation for the many-electron scenario}
As mentioned in the main text, in the many-electron scenario we have that the initial state of the system corresponds to a completely filled Fermi sea, i.e., all Wannier sites are occupied. However, we work under the assumption that different electrons barely interact with each other, such that they can be treated independently. Thus, after introducing the corresponding unitary transformations as done in the single-electron analysis, and neglecting the contributions of the conduction band, the Schrödinger equation reads
\begin{equation}
    i\hbar \dv{\tPsi}{t}
        \approx e \sum_{\vb{m}} \Tilde{\vb{R}}(t)\cdot\hat{\vb{E}}(t) \big\langle w_{v,\vb{m}}\tPsi\ket{w_{v,\vb{m}}}, 
\end{equation}
where $\ket{w_{v,\vb{m}}} \equiv \ket{w_{v,m_0}}\otimes\ket{w_{v,m_1}}\otimes \cdots \otimes \lvert w_{v,m_{N}}\rangle$; and $\Tilde{\vb{R}} = \sum_{j=0}^{N-1} \hat{\vb{R}}_j$, with $N$ the number of electrons that contribute in a phase-matched way to the HHG process.

Similarly to what we did in the single-electron analysis, we consider solid systems for which the HHG process is not much delocalized, such that the single-electron wavepacket does not spread too much over different Wannier sites. Thus, we split the $\Tilde{X}(t)$ contribution in diagonal and off-diagonal elements such that the latter can be treated using perturbation theory. Thus, by projecting our equation with respect to $\ket{w_{v,\vb{i}}}$, we get
\begin{equation}
    i\hbar \dv{\ket{\Phi_{\vb{i}}(t)}}{t}
        = e \sum_{\vb{m}} \mel{w_{v,\vb{i}}}{\Tilde{\vb{R}}(t)}{w_{v,\vb{m}}}\cdot\hat{\vb{E}}(t) \ket{\Phi_{\vb{m}}(t)},
\end{equation}
where we have defined $\ket{\Phi_{\vb{i}}(t)} = \big\langle w_{v,\vb{i}}\tPsi$, and where the matrix elements of $\Tilde{\vb{R}}(t)$ are given by
\begin{equation}
    \mel{w_{v,\vb{i}}}{\Tilde{\vb{R}}(t)}{w_{v,\vb{m}}}
        = \sum_{j=0}^{N_{\text{ph}}-1}
            \mel{w_{i_j}}{\hat{\vb{R}}_j}{w_{m_j}}
            \prod_{n \in \mathcal{O}_j} \delta_{i_n,m_n},
\end{equation}
where in the previous expression $\mathcal{O}_j:=\{n: n \in \mathbbm{N}-\{j\}\}$.

Up to the first order in perturbation theory, the equations we get are identical to the ones found in the single-electron regime. Thus, taking into account the initial conditions, we find two kinds of contributions depending on the final sites $j$ where the electrons end up in:
\begin{itemize}
    \item If the electrons end up the dynamics in the initial Wannier sites where they were initially located, then the final state reads
    \begin{equation}
        \ket{\Phi_{\vb{i}}(t)}
            = \hat{\mathcal{D}}
                \big(
                    N\boldsymbol{\chi}(t,t_0)
                \big)
                \bigotimes_q \ket{0_q};
    \end{equation}
    \item If one of the electrons ends up the dynamics in a different Wannier site $j$ from which it initially started the dynamics, then the final state reads
    \begin{align}
        \ket{\Phi_{\vb{i}}(t)}
            &= -\dfrac{i}{\hbar}
                 \hat{\mathcal{D}}
                    \big(
                        N\boldsymbol{\chi}(t,t_0)
                    \big)\nonumber
                    \\&\quad\times
                    \int^t_{t_0}
                        \dd t' e^{-i\theta(t,t',t_0)}
                            \vb{M}_{i_n,i_{n,0}}(t')\nonumber
                            \\&\hspace{1cm}\times
                            \hat{\mathcal{D}}^\dagger
                                    \big(
                                        N
                                        \boldsymbol{\chi}(t',t_0)
                                    \big)\hat{E}(t')
                               \hat{\mathcal{D}}
                                    \big(
                                        N
                                        \boldsymbol{\chi}(t',t_0)
                                    \big)\nonumber
                            \\&\hspace{1cm}
                            \bigotimes_q \ket{0_q},
    \end{align}
\end{itemize}
where we have considered that one of the electrons has transitioned from its initial site $i_{n,0}$ to site $i_{n}$. Note that the first-order perturbation theory term describes events where only one of the electrons ends up in a different Wannier site from which it initially started the dynamics. In order to consider events where we find $n$ transitions, we then have to perform a perturbation theory expansion up to the $n$th order. However, in the many-electron scenario, we restrict ourselves to solids for which the HHG process is highly localized, as it happens for instance with solid argon \cite{ndabashimiye_solid-state_2016}. In consequence, for the many-electron regime analysis we neglect the effect of intraband transitions along the valence band, and study them in the few-electron analysis.

\section*{ACKNOWLEDGEMENTS}
ICFO group acknowledges support from: ERC AdG NOQIA; Agencia Estatal de Investigación (R$\&$D project CEX2019-000910-S, funded by MCIN/ AEI/10.13039/501100011033, Plan National FIDEUA PID2019-106901GB-I00, FPI, QUANTERA MAQS PCI2019-111828-2, QUANTERA DYNAMITE PCI2022-132919, Proyectos de I+D+I “Retos Colaboración” QUSPIN RTC2019-007196-7); Fundació Cellex; Fundació Mir-Puig; Generalitat de Catalunya through the European Social Fund FEDER and CERCA program (AGAUR Grant No. 2017 SGR 134, QuantumCAT \ U16-011424, co-funded by ERDF Operational Program of Catalonia 2014-2020); the computer resources and technical support at Barcelona Supercomputing Center MareNostrum (FI-2022-1-0042); EU Horizon 2020 FET-OPEN OPTOlogic (Grant No 899794); National Science Centre, Poland (Symfonia Grant No. 2016/20/W/ST4/00314); European Union’s Horizon 2020 research and innovation programme under the Marie-Skłodowska-Curie grant agreement No 101029393 (STREDCH) and No 847648 (“La Caixa” Junior Leaders fellowships ID100010434: LCF/BQ/PI19/11690013, LCF/BQ/PI20/11760031, LCF/BQ/PR20/11770012, LCF/BQ/PR21/11840013); the Government of Spain (FIS2020-TRANQI and Severo Ochoa CEX2019-000910-S).

P. Tzallas group at FORTH acknowledges LASERLABEUROPE V (H2020-EU.1.4.1.2 grant no.871124), FORTH Synergy Grant AgiIDA (grand no. 00133), the H2020 framework program for research and innovation under the NEP-Europe-Pilot project (no. 101007417). ELI-ALPS is supported by the European Union and co-financed by the European Regional Development Fund (GINOP Grant No. 2.3.6-15-2015-00001).

J.R-D. acknowledges support from the Secretaria d'Universitats i Recerca del Departament d'Empresa i Coneixement de la Generalitat de Catalunya, as well as the European Social Fund (L'FSE inverteix en el teu futur)--FEDER. 
P.S. acknowledges funding from the European Union’s Horizon 2020 research and innovation programme under the Marie Skłodowska-Curie grant agreement No 847517. 
A.S.M. acknowledges funding support from the European Union’s Horizon 2020 research and innovation programme under the Marie Sk\l odowska-Curie grant agreement, SSFI No.\ 887153.
E.P. acknowledges the Royal Society for University Research Fellowship funding under URF$\setminus$R1$\setminus$211390.
M.F.C. acknowledges financial support from the Guangdong Province Science and Technology Major Project (Future functional materials under extreme conditions - 2021B0301030005).
\end{document}


\allowdisplaybreaks

\title{Supplementary Material of \emph{Quantum optical engineering in strong-laser driven semiconductors: A protocol for developing non-classical light sources}}

\author{J.~Rivera-Dean}
\email{javier.rivera@icfo.eu}
\affiliation{ICFO -- Institut de Ciencies Fotoniques, The Barcelona Institute of Science and Technology, 08860 Castelldefels (Barcelona)}

\author{P. Stammer}
\affiliation{ICFO -- Institut de Ciencies Fotoniques, The Barcelona Institute of Science and Technology, 08860 Castelldefels (Barcelona)}

\author{A. S. Maxwell}
\affiliation{Department of Physics and Astronomy, Aarhus University, DK-8000 Aarhus C, Denmark}

\author{Th. Lamprou}
\affiliation{Foundation for Research and Technology-Hellas, Institute of Electronic Structure \& Laser, GR-70013 Heraklion (Crete), Greece}
\affiliation{Department of Physics, University of Crete, P.O. Box 2208, GR-70013 Heraklion (Crete), Greece}

\author{A. F. Ordóñez}
\affiliation{ICFO -- Institut de Ciencies Fotoniques, The Barcelona Institute of Science and Technology, 08860 Castelldefels (Barcelona)}

\author{E. Pisanty}
\affiliation{Department of Physics, King's College London, WC2R 2LS London, United Kingdom}

\author{P. Tzallas}
\affiliation{Foundation for Research and Technology-Hellas, Institute of Electronic Structure \& Laser, GR-70013 Heraklion (Crete), Greece}
\affiliation{ELI-ALPS, ELI-Hu Non-Profit Ltd., Dugonics tér 13, H-6720 Szeged, Hungary}

\author{M. Lewenstein}
\email{maciej.lewenstein@icfo.eu}
\affiliation{ICFO -- Institut de Ciencies Fotoniques, The Barcelona Institute of Science and Technology, 08860 Castelldefels (Barcelona)}
\affiliation{ICREA, Pg. Llu\'{\i}s Companys 23, 08010 Barcelona, Spain}

\author{M. F. Ciappina}
\email{marcelo.ciappina@gtiit.edu.cn}
\affiliation{Physics Program, Guangdong Technion--Israel Institute of Technology, Shantou, Guangdong 515063, China}
\affiliation{Technion -- Israel Institute of Technology, Haifa, 32000, Israel}
\affiliation{Guangdong Provincial Key Laboratory of Materials and Technologies for Energy Conversion, Guangdong Technion – Israel Institute of Technology, Shantou, Guangdong 515063, China}

\maketitle
\tableofcontents
\clearpage
\section{Wannier and Bloch states}\label{App:Wannier:Bloch}
According to Bloch's theorem, the eigenstates of a single-electron Hamiltonian with a periodic potential, can be written as the product of a plane wave and a function that has the periodicity of the applied potential \cite{Ashcroft_book}. In particular, for a one-dimensional system the Bloch functions read
\begin{equation}
    \braket{x}{\phi_{m,k}}
        =\phi_{m,k}(x)
        = u_{m,k}(x)e^{ikx},
\end{equation}
where $m$ is a subscript specifying the band over which the corresponding Bloch wavefunction is defined, and $k$ is the crystal momentum. Another usual representation of the quantum state of an electron within the solid is by means of the so-called Wannier functions, which are given by
\begin{equation}
    \braket{x}{w_{m,j}}
        =w_{m,j}(x)
        =\int_{\text{BZ}} 
            \dd k \ \phi_{m,k}(x-x_j)
            \Tilde{w}_m(k),
\end{equation}
where $\Tilde{w}_m(k)$ is a product of a normalization constant and a phase that depends on the crystal momentum $k$. However, and as shown in Ref.~\cite{kohn_analytic_1959}, for a 1D lattice this function is independent of $k$. We observe that, unlike the set of Bloch wavefunctions which are completely delocalized in space due to their plane wave nature, Wannier states are defined as a superposition within the Brillouin zone of all possible Bloch states, which leads a localized wavepacket within the real space \cite{marzari_maximally_2012}.

\subsection{Computing the matrix element $\mel{w_{v,j}}{\hat{X}}{\phi_{c,k}}$}
In order to compute the different quantities we study along the main text, we need to use the previous definitions of Bloch and Wannier states. Of particular importance for our purposes, is the transition matrix element between a Wannier-valence band state and a Bloch-conduction band state, i.e., $d_{j,c}(k)=e\mel{w_{v,j}}{\hat{X}}{\phi_{c,k}}$. First, we introduce the identity in the position representation so that the previous element can be written as
\begin{equation}
    \begin{aligned}
    d_{j,c}(k)
        &= e \int \dd x \ w^*_{v,j}(x) x \phi_{c,k}(x)
        \\&
        = e\Tilde{w}^*_v\int \dd x
            \int_{\text{BZ}} \dd k'
                \phi_{v,k'}^*(x-x_j) x \phi_{c,k}(x).
    \end{aligned}
\end{equation}

Noting that Bloch wavefunctions are periodic, so that for site $x_j = j \mathsf{a}$, with $\mathsf{a}$ the lattice constant, we can write
\begin{equation}\label{Eq:App:rel:Bloch}
    \phi_{c,k}(x-x_j)
        = \phi_{c,k}(x)e^{-ikx_j},
\end{equation}
which allows us to express the matrix element $d_{j,c}(k)$ as
\begin{equation}\label{Eq:App:djc:int}
    \begin{aligned}
    d_{j,c}(k) 
        &= e\Tilde{w}^*_v
       \int \dd x \int_{\text{BZ}}\dd k'
        \big[ 
           \phi_{v,k'}^*(x-x_j)
        \big]
                (x-x_j)
                  \big[  \phi_{c,k}(x-x_j)
                 \big]
                 e^{ikx_j}
        \\&
        = e \Tilde{w}_{v}^*e^{ikx_j}
            \int_{\text{BZ}} \dd k' 
                \mel{\phi_{v,k'}}{\hat{X}}{\phi_{c,k}},
    \end{aligned}
\end{equation}
where in going from the first to the second equality in \eqref{Eq:App:djc:int} we have taken into account the orthonormality condition between Bloch states. Then, provided that
\begin{equation}
    \begin{aligned}
    \mel{\phi_{v,k'}}{\hat{X}}{\phi_{c,k}}
        &= i\dfrac{2\pi}{\mathsf{a}} \pdv{\delta(k'-k)}{k}
            \int^{\mathsf{a}/2}_{-\mathsf{a}/2}
                \dd x \ \phi_{v,k'}(x)\phi_{c,k}(x)
        \\&\quad
        + \dfrac{2\pi}{\mathsf{a}}\delta(k'-k)
            \int^{\mathsf{a}/2}_{-\mathsf{a}/2}
                \dd x \ \phi_{v,k'}(x)x\phi_{c,k}(x),
    \end{aligned}
\end{equation}
we can then write
\begin{equation}
    d_{j,c}(k) = \Tilde{w}^*_ve^{ikx_j}d_{v,c}(k),
\end{equation}
where $d_{v,c}(k) = e\mel{\phi_{v,k}}{\hat{X}}{\phi_{c,k}}$.

\subsection{Computing the matrix element $\mel{\phi_{v,k}}{\hat{X}}{\phi_{c,k}}$}
From what we have seen above, the transition matrix element between a Wannier-valence band and a Bloch-conduction band state, depends on the transition matrix element between two Bloch states located in the corresponding bands. Hence, we are now interested in computing $d_{v,c}(k)$. With that purpose, we take into account that
\begin{equation}
    \comm{\hat{X}}{\hat{H}_{\text{cr}}}   = \dfrac{i\hbar}{m}\hat{P}_x,
\end{equation}
where $m$ is the electron mass, $\hat{H}_{\text{cr}}$ the crystal Hamiltonian, and $\hat{P}_x$ the momentum operator along the $x$ direction. According to this expression, we can write
\begin{equation}
    \mel{\phi_{v,k}}{\hat{X}}{\phi_{c,k}}
        = \dfrac{i\hbar}{m}\dfrac{ \mel{\phi_{v,k}}{\hat{P}_x}{\phi_{c,k}}}{\mathcal{E}(k)},
\end{equation}
where we have denoted $\mathcal{E}(k) = \mathcal{E}_c(k)-\mathcal{E}_v(k)$. Similarly to what we did before, we now proceed to evaluate the matrix element of the momentum operator $\hat{P}_x$
\begin{equation}
    \mel{\phi_{v,k}}{\hat{P}_x}{\phi_{c,k}}
        = -i\hbar \int^{\infty}_{-\infty} \dd x \ \phi^*_{v,k}(x)
            \pdv{\phi_{c,k}(x)}{x},
\end{equation}
and expanding the Bloch wavefunctions up to the leading term in the $\vb{k}\cdot\vb{p}$ theory \cite{HaugBook}
\begin{equation}\label{Eq:App:1st:kp}
    \phi_{m,k}(x)\simeq e^{ikx}\dfrac{u_{m,0}(x)}{\mathsf{a}^{1/2}},
\end{equation}
such that
\begin{equation}
    \pdv{\phi_{m,k}(x)}{x}
        = i k \phi_{m,k}(x) + \dfrac{e^{ikx}}{\mathsf{a}^{1/2}}
        \pdv{u_{m,0}(x)}{x},
\end{equation}
we can write the matrix element of the momentum operator as
\begin{equation}
    \mel{\phi_{v,k}}{\hat{P}_x}{\phi_{c,k}}
        = \dfrac{i\hbar}{\mathsf{a}}
            \int^{\infty}_{-\infty} \dd x \ u^*_{v,0}(x)u'_{c,0}(x) 
            \equiv p_{v,c}(0),
\end{equation}
where we have taken into account that $\braket{\phi_{v,k}}{\phi_{c,k}} = 0$. With all the above, we get for the initial transition matrix element
\begin{equation}
    d_{v,c}(k)
        = \dfrac{i\hbar e}{m}\dfrac{p_{v,c}(0)}{\mathcal{E}(k)}.
\end{equation}

In the bibliography \cite{vurgaftman_band_2001,ozgur_comprehensive_2005,schleife_band-structure_2009,yan_strain_2012}, one can find tabulated values for the so-called Kane parameter $\mathfrak{E}_{p,i}$, with $i\in\{x,y,z\}$, of different solids, which is related to the element $p_{v,c}(0)$ by $\mathfrak{E}_{p,x} = 2p_{v,c}^2(0)$. Thus, in terms of the Kane parameter, the transition matrix element reads
\begin{equation}
    d_{v,c}(k)
        = \dfrac{i\hbar e}{m}\sqrt{\dfrac{\mathfrak{E}_{p,x}}{2\mathcal{E}^2(k)}}.
\end{equation}

\subsection{Computing the matrix element $\mel{w_{v,j}}{\hat{X}}{w_{v,j'}}$}
Finally, and to conclude with this section, we proceed to compute the matrix  element $\mel{w_{v,j}}{\hat{X}}{w_{v,j'}}$. In this case, we have in the position representation
\begin{equation}
    \begin{aligned}
    \mel{w_{v,j}}{\hat{X}}{w_{v,j'}}
        &= \int \dd x w_{v,j}(x) x w_{v,j'}(x)
        \\
        &= \abs{\Tilde{w}_v}^2
            \int \dd x \ x
                \int_{\text{BZ}}\dd k
                    \int_{\text{BZ}}\dd k'
                        \phi^*_{v,k}(x-x_j)
                            \phi_{v,k'}(x-x_{j'}),
    \end{aligned}
\end{equation}
where in the last step we have introduced the definition of Wannier states in terms of the Bloch functions. By taking into account Eq.~\eqref{Eq:App:rel:Bloch}, we then have
\begin{equation}
    \begin{aligned}
    \mel{w_{v,j}}{\hat{X}}{w_{v,j'}}
        &= \abs{\Tilde{w}_v}^2
            \int_{\text{BZ}}\dd k
                \int_{\text{BZ}}\dd k'
                    e^{ikx_j-ik'x_{j'}}
                    \int \dd x 
                        \phi^*_{v,k}(x)
                            x
                            \phi_{v,k'}(x)
        \\
        &= i\abs{\Tilde{w}_v}^2
            \int_{\text{BZ}}\dd k
                \int_{\text{BZ}}\dd k'
                    e^{ikx_j-ik'x_{j'}}
                    \pdv{}{k}\delta(k-k')
        = \abs{\Tilde{w}_v}^2 \dfrac{2\pi}{\mathsf{a}} x_{j'} \delta_{j,j'},
    \end{aligned}
\end{equation}
where in going from the first equality to the second equality we have considered up to the leading term in the $\vb{k}\cdot \vb{p}$ theory (see Eq.~\eqref{Eq:App:1st:kp}). Furthermore, we have taken into account that when applying normalization conditions over Wannier functions in 1D, we can write $\abs{\Tilde{w}_v}^2 = \mathsf{a}/(2\pi)$.

\section{Quantum state of the system after HHG}
After high-harmonic generation processes, and neglecting the conduction band contributions, we found in the main text that the state of the system is given by
\begin{equation}\label{Eq:noncond:state}
    \ket{\Psi(t)}
        = D(\alpha)
            \bigg[
                \ket{w_{v,j_0}}\ket{\Phi_{j_0}(t)}
                + \sum_{j\neq j_0} \ket{w_{v,j}}\ket{\Phi_{j}(t)}
            \bigg],
\end{equation}
where we have that each of the quantum optical components $\{\ket{\Phi_{j}(t)}:j\in\mathbbm{Z}\}$ are given by
\begin{align}
    &\ket{\Phi_{j_0}(t)}
        = \bigg(
            \prod_q
                e^{i\varphi_q(t,t_0)}
                \hat{D}
                    \big(
                        \chi_q(t,t_0)
                    \big)
          \bigg)
        \ket{\bar{0}},
    \label{Eq:noncond:j0}
    \\&
    \ket{\Phi_{j}(t)}
        = \dfrac{i}{\hbar}
        \bigg(
            e^{i\varphi_q(t,t_0)}
            \prod_q \hat{D}
                \big(
                    \chi_q(t,t_0)
                \big)
        \bigg)
            \int^{t}_{t_0}
                \dd t' e^{-i\theta(t,t',t_0)} M_{j,j_0}(t')
                    \big[
                        \hat{E}(t') + \Tilde{E}_{\text{cl}}(t')
                    \big]
                    \ket{\bar{0}}
    \label{Eq:noncond:j},
\end{align}
where in these expressions we have denoted the vacuum state of all modes $\ket{\bar{0}} = \bigotimes_q \ket{0_q}$, $\theta(t,t',t_0) = \text{Im}[\chi(t,t_0)\chi^*(t,t_0)]$ and $\Tilde{E}_{\text{cl}}(t) = \mel{\boldsymbol{\chi}(t',t_0)}{\hat{E}(t')}{\boldsymbol{\chi}(t',t_0)}$, with $\ket{\boldsymbol{\chi}(t',t_0)} = \ket{\chi_1(t',t_0)} \otimes \cdots \otimes \ket{\chi_N(t',t_0)}$. Here, we are interested in observing some of the possible non-classical properties of the light state obtained from Eq.~\eqref{Eq:noncond:state}. Thus, as we have no knowledge about in which state the electron may have recombined, we trace out the electronic degrees of freedom to get
\begin{equation}
    \rho(t)
        = D(\alpha) \dyad{\Phi_{j_0}(t)}D^\dagger(\alpha)
            + \sum_{j\neq j_0}
                D(\alpha)\dyad{\Phi_{j}(t)}D^\dagger(\alpha).
\end{equation}

In a similar basis to what was studied in Refs.~\cite{lewenstein_generation_2021,rivera-dean_strong_2022}, here we are going to study the Wigner function representation of the fundamental mode and, with that purpose, we project the state of the harmonics into some coherent state $\ket{\boldsymbol{\gamma}}$
\begin{equation}
    \begin{aligned}
    \rho_{q=1,j}(t)
        &= \mel{\boldsymbol{\gamma}}{\rho(t)}{\boldsymbol{\gamma}}
        = D(\alpha) \braket{\boldsymbol{\gamma}}{\Phi_{j_0}(t)}
        \braket{\Phi_{j_0}(t)}{\boldsymbol{\gamma}}
        D^\dagger(\alpha)
            + \sum_{j\neq j_0}
                D(\alpha) \braket{\boldsymbol{\gamma}}{\Phi_{j}(t)}
            \braket{\Phi_{j}(t)}{\boldsymbol{\gamma}}
            D^\dagger(\alpha),
    \end{aligned}
\end{equation}
where we have that
\begin{align}
    &\braket{\boldsymbol{\gamma}}{\Phi_{j_0}(t)}
    = \braket{\boldsymbol{\gamma}}{\boldsymbol{\chi}_{q\neq 1}(t,t_0)}
    \ket{\chi_1(t,t_0)},
    \\&\braket{\boldsymbol{\gamma}}{\Phi_{j}(t)}
        = \dfrac{i}{\hbar}
        \int^t_{t_0}\dd t' e^{-i\theta(t,t',t_0)}
        M_{j,j_0}(t')
        \Big[
            -i g(\omega_1)e^{i\omega_1 t'} \braket{\boldsymbol{\gamma}}{\boldsymbol{\chi}_{q\neq 1}(t,t_0)}
            D\big(\chi_1(t,t_0)\big)\ket{1_1}\nonumber
            \\&\hspace{6.25cm}
            -i \sum_{q\neq 1}g(\omega_q)
            e^{i\omega_q t'}
            \mel{\boldsymbol{\gamma}}{D\big(\boldsymbol{\chi}_{q\neq 1}(t,t_0)\big)}{1_q,\bar{0}_{q'\neq (q,1)}}
            \ket{\chi_1(t,t_0)}\nonumber
            \\&\hspace{6.25cm}
            + \Tilde{E}_{\text{cl}}(t')
                \braket{\boldsymbol{\gamma}}{\boldsymbol{\chi}_{q\neq 1}(t,t_0)} 
                \ket{\chi_1(t,t_0)}
        \Big]
        \\&\hspace{1.5cm}
        =
        h_1(t,t_0) D\big(\chi_1(t,t_0)\big)\ket{1_1}
        + h_2(t,t_0) \ket{\chi_1(t,t_0)},
\end{align}
where in this last expression we have defined
\begin{align}
    &h_1(t,t_0)
        = g(\omega_1)\dfrac{1}{\hbar}
        \braket{\boldsymbol{\gamma}}{\boldsymbol{\chi}_{q\neq 1}(t,t_0)}
        \int^t_{t_0}\dd t' e^{-i\theta(t,t',t_0)}
        M_{j,j_0}(t')e^{i\omega_1 t},
    \\&
    h_2(t,t_0)
        =\dfrac{i}{\hbar}
        \int^t_{t_0}\dd t' e^{-i\theta(t,t',t_0)}
        M_{j,j_0}(t')
        \Big[
            -i \sum_{q\neq 1}g(\omega_q)
            e^{i\omega_q t'}
            \mel{\boldsymbol{\gamma}}{D\big(\boldsymbol{\chi}_{q\neq 1}(t,t_0)\big)}{1_q,\bar{0}_{q'\neq (q,1)}}
            + \Tilde{E}_{\text{cl}}(t')
                \braket{\boldsymbol{\gamma}}{\boldsymbol{\chi}_{q\neq 1}(t,t_0)}
        \Big].
\end{align}

\subsection{Wigner function of the fundamental mode}
Following A.~Royer's approach \cite{royer_wigner_1977}, we can compute the Wigner function of this state as
\begin{equation}
    W(\beta,t)
        = \dfrac{2}{\mathcal{N}\pi}
            \tr(D(\beta)\Pi D^\dagger(\beta)\rho_{q=1}(t)),
\end{equation}
where $\mathcal{N} = \braket{\boldsymbol{\gamma}}{\boldsymbol{\chi}_{q\neq 1}(t,t_0)} + \abs{h_1(t,t_0)}^2 + \abs{h_2(t,t_0)}^2$. In our case, the Wigner function can be written as a sum of two components
\begin{equation}
    W(\beta) = W_{j_0}(\beta)
                + \sum_{j\neq j_0} W_{j}(\beta), 
\end{equation}
where we have that each of the terms in the above expression are given by
\begin{align}
    &W_{j_0}(\beta)
        = \braket{\boldsymbol{\gamma}}{\boldsymbol{\chi}_{q\neq 1}(t,t_0)} e^{-2\abs{\beta}^2},
    \\
    &W_{j}(\beta)
        =\big[
            \abs{h_1(t,t_0)}^2
                (4\abs{\beta}^2-1) 
            + \abs{h_2(t,t_0)}^2
            + h_1(t,t_0)^*h_2(t,t_0)
                2\beta
            + h_2(t,t_0)^*h_1(t,t_0)
                2\beta^*
        \big]e^{-2\abs{\beta}^2}
\end{align}
where we have set our origin in phase space to $\beta_0 = \alpha + \chi_1(t,t_0)$.

\section{Quantum state of the system after the \emph{conditioning to HHG} operation}
In this section, we introduce the so-called \emph{conditioning to HHG} operations \cite{lewenstein_generation_2021,rivera-dean_strong_2022,stammer_high_2022,stammer_theory_2022,stammer_quantum_2022,rivera-dean_light-matter_2022} in order to further explot the non-classical properties of the generated state. In short, this operation consists of looking at those events in which at least one excitation in one of the field modes is generated, but taking into account the correlations that are established with the fundamental mode. Mathematically speaking, we consider the projective operation
\begin{equation}
    \hat{P}_{\text{HHG}}
        = \mathbbm{1} - \dyad{\alpha}\bigotimes_{q>1} \dyad{0_q},
\end{equation}
i.e., we look at all those events where both fundamental and harmonic modes are not found in the original state in which they initially started the dynamics. By implementing this operation to the state in Eq.~\eqref{Eq:noncond:state} we get
\begin{equation}
    \begin{aligned}
    \ket{\Psi(t)}
            &= D(\alpha)
                \bigg[
                    \ket{w_{v,j_0}}
                    \Big(
                        \ket{\Phi_{j_0}(t)} 
                        - \braket{\bar{0}}{\Phi_{j_0}(t)}
                            \ket{\bar{0}}
                    \Big) 
                    + \sum_{j\neq j_0} \ket{w_{v,j}}
                    \Big(
                        \ket{\Phi_{j}(t)} 
                        - \braket{\bar{0}}{\Phi_j(t)}
                            \ket{\bar{0}}
                    \Big)
                \bigg]
            \\
            &\equiv D(\alpha)
            \bigg[
                \ket{w_{v,j_0}}\ket{\bar{\Phi}_{j_0}(t)}
                + \sum_{j\neq j_0} \ket{w_{v,j}}\ket{\bar{\Phi}_{j}(t)}
            \bigg],
    \end{aligned}
\end{equation}
where we have defined
\begin{align}
    &\ket{\bar{\Phi}_{j_0}(t)}
        = \ket{\Phi_{j_0}(t)} 
            - \braket{\bar{0}}{\Phi_{j_0}(t)}
                            \ket{\bar{0}},
    \quad \quad
    \ket{\bar{\Phi}_{j}(t)}
        = \ket{\Phi_{j}(t)} 
            - \braket{\bar{0}}{\Phi_j(t)}
                            \ket{\bar{0}}.
\end{align}

By explicitly introducing Eqs.~\eqref{Eq:noncond:j0} and Eq.~\eqref{Eq:noncond:j} in the expressions above, we respectively find
\begin{align}
    &\ket{\bar{\Phi}_{j_0}(t)}
        = e^{i\sum_q\varphi_q(t,t_0)}
        \big(
            \ket{\boldsymbol{\chi}(t,t_0)}
            - \braket{\bar{0}}{\boldsymbol{\chi}(t,t_0)}
                \ket{\bar{0}}
        \big),\label{Eq:cond:j0}
    \\&
    \ket{\bar{\Phi}_{j}(t)}
        =e^{i\sum_q\varphi_q(t,t_0)}
        \dfrac{i}{\hbar}
            \int^t_{t_0} \dd t' 
                e^{i\theta(t,t',t_0)}M_{j,j_0}(t')\nonumber
    \\&\hspace{4.5cm}\times
            \bigg\{
                -i\sum_q g(\omega_q)e^{i\omega_q t'}
                \Big[
                    \hat{D}
                        \big(
                            \boldsymbol{\chi}(t,t_0)
                        \big)
                        \ket{1_q,\bar{0}_{q'\neq q}}
                    - \braket{-\boldsymbol{\chi}(t',t_0)}{1_q,\bar{0}_{q'\neq q}}
                        \ket{\bar{0}}
                \Big]
                \nonumber
                \\
                &\hspace{5.3cm}
                + \Tilde{E}_{\text{cl}}(t')
                    \Big[
                        \ket{\boldsymbol{\chi}(t,t_0)}
                        -\braket{\bar{0}}{\boldsymbol{\chi}(t,t_0)}\ket{\bar{0}}
                    \Big]
            \bigg\},\label{Eq:cond:j}
\end{align}
and we observe that the two contributions are written in terms of superpositions between displaced Fock states, which depending on the amount of displacement and the contribution of the $j\neq j_0$ terms, could potentially lead to the generation of non-classical light states. Note that although the expressions we have here are the ones for the single-electron scenario, can be trivially extended to the multi-electron scenario by multipying the displacement by the number of phase-matched electrons contributing to HHG, as explained in the Methods section.

\subsection{Wigner function of the fundamental mode}
In the conditioning operation, the state of the harmonic modes is typically measured in some coherent state $\ket{\boldsymbol{\gamma}}\equiv \bigotimes_{q>1} \ket{\gamma_q}$. Here, we consider such operation and project the states in Eqs.~\eqref{Eq:cond:j0} and \eqref{Eq:cond:j} with respect to this coherent state
\begin{align}
    &\braket{\boldsymbol{\gamma}}{\bar{\Phi}_{j_0}(t)}
        = e^{i\sum_q\varphi_q(t,t_0)}
        \big(
            \braket{\boldsymbol{\gamma}}{\boldsymbol{\chi}_{q\neq1}(t,t_0)}\ket{\chi_1(t,t_0)}
            - \braket{\bar{0}}{\boldsymbol{\chi}(t,t_0)}\braket{\boldsymbol{\gamma}}{\bar{0}_{q\neq 1}}\ket{0_1}
        \big),
    \\&
    \braket{\boldsymbol{\gamma}}{\bar{\Phi}_{j}(t)}
        =e^{i\sum_q\varphi_q(t,t_0)}
        \dfrac{i}{\hbar}
            \int^t_{t_0} \dd t' 
                e^{i\theta(t,t',t_0)}M_{j,j_0}(t')\nonumber
    \\&\hspace{4.5cm}\times
            \bigg\{
                -i g(\omega_1)e^{i\omega_1 t'}
                \braket{\boldsymbol{\gamma}}{\boldsymbol{\chi}_{q\neq 1}(t,t_0)}
                    \hat{D}
                        \big(
                            \chi_1(t,t_0)
                        \big)
                        \ket{1_q}
                \\&\hspace{5.2cm}
                    -i \sum_{q>1} g(\omega_q)e^{i\omega_q t'}
                \mel{\boldsymbol{\gamma}}{\hat{D}(\boldsymbol{\chi})}{1_q,\bar{0}_{q\neq (q,1)}}\ket{0_1}\nonumber
                \\&\hspace{5.2cm}
                +i\sum_{q}g(\omega_q)
                    e^{i\omega_q t'}
                    \braket{-\boldsymbol{\chi}(t,t_0)}{1_q,\bar{0}_{q'\neq q}}
                    \braket{\boldsymbol{\gamma}}{\bar{0}_{q'\neq 1}}
                        \ket{\bar{0}_1}
                \nonumber
                \\
                &\hspace{5.3cm}
                + \Tilde{E}_{\text{cl}}(t')
                    \Big[
                        \braket{\boldsymbol{\gamma}}{\boldsymbol{\chi}_{q\neq1}(t,t_0)}\ket{\chi_1(t,t_0)}
                        - \braket{\bar{0}}{\boldsymbol{\chi}(t,t_0)}\braket{\boldsymbol{\gamma}}{\bar{0}_{q\neq 1}}\ket{0_1}
                    \Big]
            \bigg\},
\end{align}
and for the sake of clarity, we rewrite the last expression as
\begin{equation}
    \braket{\boldsymbol{\gamma}}{\bar{\Phi}_{j}(t)}
        = \bar{h}_1(t,t_0) \hat{D}\big(\chi_1(t,t_0)\big)\ket{1_1}
        + \bar{h}_2(t,t_0)
        \ket{\chi_1(t,t_0)}
        + \bar{h}_3(t,t_0)\ket{0_1},
\end{equation}
where we have defined
\begin{align}
    &\bar{h}_1(t,t_0)
        = \dfrac{1}{\hbar}
             \braket{\boldsymbol{\gamma}}{\boldsymbol{\chi}_{q\neq 1}(t,t_0)}
            \int^t_{t_0} \dd t' 
                e^{i\theta(t,t',t_0)}M_{j,j_0}(t')
                g(\omega_1)e^{i\omega_1 t'}
    \\
    &\bar{h}_2(t,t_0)
        = \dfrac{i}{\hbar}
            \int^t_{t_0} \dd t' 
                e^{i\theta(t,t',t_0)}M_{j,j_0}(t')
                \bigg[
                    -i \sum_{q>1} g(\omega_q)e^{i\omega_q t'}
                \mel{\boldsymbol{\gamma}}{\hat{D}(\boldsymbol{\chi}_{q\neq 1}(t,t_0))}{1_q,\bar{0}_{q\neq (q,1)}}
                + \Tilde{E}_{\text{cl}}(t')
                        \braket{\boldsymbol{\gamma}}{\boldsymbol{\chi}_{q\neq1}(t,t_0)}
                \bigg],
    \\
    &\bar{h}_3(t,t_0)
        = \dfrac{i}{\hbar}
            \int^t_{t_0} \dd t' 
                e^{i\theta(t,t',t_0)}M_{j,j_0}(t')
                \bigg[
                    i\sum_{q}g(\omega_q)
                    e^{i\omega_q t'}
                    \braket{-\boldsymbol{\chi}(t,t_0)}{1_q,\bar{0}_{q'\neq q}}
                    \braket{\boldsymbol{\gamma}}{\bar{0}_{q'\neq 1}}
                    - \Tilde{E}_{\text{cl}}(t')
                        \braket{\bar{0}}{\boldsymbol{\chi}(t,t_0)}\braket{\boldsymbol{\gamma}}{\bar{0}_{q\neq 1}}
                \bigg].
\end{align}

With all the above, we can write the quantum optical state of the fundamental mode after applying the conditioning operation as
\begin{equation}
    \bar{\rho}_1(t)
        = \braket{\boldsymbol{\gamma}}{\bar{\Phi}_{j_0}(t)}\braket{\bar{\Phi}_{j_0}(t)}{\boldsymbol{\gamma}}
        + \sum_{j\neq j_0}
            \braket{\boldsymbol{\gamma}}{\bar{\Phi}_{j}(t)}\braket{\bar{\Phi}_{j}(t)}{\boldsymbol{\gamma}},
\end{equation}
and the associated Wigner function can be written as
\begin{equation}
    \widetilde{W}(\beta)
        = \dfrac{2}{\pi \mathcal{N}}
        \Big(
            \widetilde{W}_{j_0}(\beta)
            +   \sum_{j\neq j_0}
            \widetilde{W}_{j}(\beta)
        \Big),
\end{equation}
where we have that
\begin{align}
    &\widetilde{W_{j_0}}(\beta)
        = \abs{\braket{\boldsymbol{\gamma}}{\boldsymbol{\chi}_{q\neq1}(t,t_0)}}^2
        e^{-2\abs{\beta-\chi_1(t,t_0)}^2}
        + \abs{\braket{\bar{0}}{\boldsymbol{\chi}(t,t_0)}\braket{\boldsymbol{\gamma}}{\bar{0}_{q\neq 1}}}^2
            e^{-2\abs{\beta}^2}\nonumber
        \\& \hspace{1.5cm}
        -
            \braket{\boldsymbol{\gamma}}{\boldsymbol{\chi}_{q\neq1}(t,t_0)}
            \braket{\bar{0}}{\boldsymbol{\chi}(t,t_0)}^*\braket{\boldsymbol{\gamma}}{\bar{0}_{q\neq 1}}^*
            e^{i2\Im(\chi_1(t,t_0)\beta^*)}
            e^{-\frac{1}{2}\abs{2\beta - \chi_1(t,t_0)}^2}\nonumber
        \\& \hspace{1.5cm}
        - 
        \braket{\boldsymbol{\gamma}}{\boldsymbol{\chi}_{q\neq1}(t,t_0)}^*
            \braket{\bar{0}}{\boldsymbol{\chi}(t,t_0)}\braket{\boldsymbol{\gamma}}{\bar{0}_{q\neq 1}}
            e^{-i2\Im(\chi_1(t,t_0)\beta^*)}
            e^{-\frac{1}{2}\abs{2\beta - \chi_1(t,t_0)}^2},
    \\& \widetilde{W}_{j}(\beta)
        = \abs{\bar{h}_1(t,t_0)}^2(4\abs{\beta-\chi_1(t,t_0)}^2-1)e^{-2\abs{\beta - \chi_1(t,t_0)}^2}
        + \abs{\bar{h}_2(t,t_0)}^2
        e^{-2\abs{\beta - \chi_1(t,t_0)}^2}
        +\abs{\bar{h}_3(t,t_0)}^2
        e^{-2\abs{\beta}^2}\nonumber
        \\& \hspace{1.5cm}
        +\big[
            2(\beta-\chi_1(t,t_0))^*
            \bar{h}_1(t,t_0)^*\bar{h}_2(t,t_0)
            + 2(\beta-\chi_1(t,t_0))
            \bar{h}_1(t,t_0)\bar{h}_2(t,t_0)^*
        \big]
            e^{-2\abs{\beta - \chi_1(t,t_0)}^2}\nonumber
    \\&\hspace{1.5cm}
        + \big[
            e^{-i2\Im(\chi_1(t,t_0)\beta^*)}
            (2\beta-\chi_1(t,t_0))
            \bar{h}_1(t,t_0)^*\bar{h}_3(t,t_0)\nonumber
            \\&\hspace{2cm}
            + 
            e^{i2\Im(\chi_1(t,t_0)\beta^*)}
            (2\beta-\chi_1(t,t_0))^*
            \bar{h}_1(t,t_0)\bar{h}_3(t,t_0)^*
        \big]
        e^{-\frac12\abs{2\beta-\chi_1(t,t_0)}^2}\nonumber
        \\&\hspace{1.5cm}
        +\big[
            e^{-i2\Im(\chi_1(t,t_0)\beta^*)}
            \bar{h}_2(t,t_0)^*\bar{h}_3(t,t_0)
            + e^{i2\Im(\chi_1(t,t_0)\beta^*)}
            \bar{h}_2(t,t_0)\bar{h}_3(t,t_0)^*
        \big]
        e^{-\frac12\abs{2\beta-\chi_1(t,t_0)}^2}.
n\end{align}

\section{Light-matter entanglement characterization}
In this section, we are going to consider the case where we have $N$ electrons contributing in a phase-matched way to the HHG process. In line with our approximations, we consider here the case where the HHG process is highly localized so most of the electrons recombine with their initial site. However, we allow one out of the $N$ electrons end up the dynamics in their nearest-neighbor sites. Thus, the state of the system is given by
\begin{equation}
    \ket{\Psi(t)}
        = D(\alpha)
            \bigg[
                \ket{w_{v,\boldsymbol{j_0}}}\ket{\Phi_{\boldsymbol{j_0}}(t)}
                + \sum_{\boldsymbol{j}\in \text{NN}}
                \ket{w_{v,\boldsymbol{j}}}
                \ket{\Phi_{\boldsymbol{j}}(t)}
            \bigg],
\end{equation}
where $\boldsymbol{j_0}$ is a vector describing the initial site of all possible electrons, while $\text{NN} = \{ \boldsymbol{j}: \boldsymbol{j}_0 \pm \boldsymbol{1}_j \ \text{with} \ \boldsymbol{1}_j = (0_1, 0_2, \cdots,0_{j-1}, 1_j,0_{j+1}, \cdots, 0_{N}) \ \forall j\}$. Since all Wannier sites are equivalent and the electrons behave independently, then no matter which electron ends up the dynamics in a nearest-neighboring site that its effect on the quantum optical state of the system will not depend on it. Thus, we can simplify our state and write it as
\begin{equation}
    \ket{\Psi(t)}
        = D(\alpha)
            \bigg[
                \ket{w_{v,\boldsymbol{j_0}}}\ket{\Phi_{\boldsymbol{j_0}}(t)}
                + \sqrt{f_c N}
                \ket{w_{v,\text{NN}}}
                \ket{\Phi_{\boldsymbol{j}}(t)},
            \bigg]
\end{equation}
where $f_c$ is the coordinate number of the lattice and where we have defined
\begin{equation}
    \ket{w_{v,\text{NN}}}
        = \dfrac{1}{\sqrt{f_c N}}
            \sum_{\boldsymbol{j}\in \text{NN}}
            \ket{w_{v,\boldsymbol{j}}}.
\end{equation}

We note that the expressions we have derived so far for the Wigner function can be trivially extended to include this multi-electron scenario. However, in this section we are going to study the entanglement properties between the electronic and quantum optical degrees of freedom. Thus, we see that for solid-state systems for which the HHG process is highly localized, we can study the entanglement properties by just considering two-electronic states $\{\ket{w_{v,\boldsymbol{j}_0}}, \ket{w_{v,\text{NN}}}\}$, which greatly simplifies the entanglement characterization by means of the entropy of entanglement.

\subsection{Entanglement characterization after HHG}
Here, we are going to study the entanglement properties of the state after HHG processes, between the fundamental electromagnetic field mode and the electronic degrees of freedom. Thus, we work with the following state
\begin{equation}
    \begin{aligned}
    \braket{\boldsymbol{\gamma}}{\Psi(t)}
        &= D(\alpha)
            \bigg[
                \ket{w_{v,\boldsymbol{j}_0}}
                \braket{\boldsymbol{\gamma}}{\Phi_{\boldsymbol{j_0}}(t)}
                + \sum_{\boldsymbol{j}\in\text{NN}}
                    \ket{w_{v,\boldsymbol{j}}}
                    \braket{\boldsymbol{\gamma}}{\Phi_{\boldsymbol{j}}(t)}
            \bigg]
        \\&
        =D(\alpha)
            \bigg[
                \ket{w_{v,\boldsymbol{j}_0}}
                \ket{\Phi_{\boldsymbol{j_0}}(\boldsymbol{\gamma},t)}
                + \sum_{\boldsymbol{j}\in\text{NN}}
                    \ket{w_{v,\boldsymbol{j}}}
                    \ket{\Phi_{\boldsymbol{j}}(\boldsymbol{\gamma},t)}
            \bigg],
    \end{aligned}
\end{equation}
where we have defined
\begin{equation}
    \ket{\Phi_{\boldsymbol{j}}(\boldsymbol{\gamma},t)}
        = \braket{\boldsymbol{\gamma}}{\Phi_{\boldsymbol{j}}(t)}.
\end{equation}

Our aim now is to compute the entropy of entanglement $S(\rho) := -\tr(\rho \log_2 \rho)$, where $\rho$ is the partial trace with respect to one of the subsystems. In our case, and for the sake of simplicity, we will consider the partial trace with respect to the quantum optical degrees of freedom such that the resulting density matrix reads
\begin{equation}
    \begin{aligned}
    \rho_{\text{elec}}(t)
        &= 
                \dyad{w_{v,\boldsymbol{j_0}}}
                \braket{\Phi_{\boldsymbol{j_0}}(\boldsymbol{\gamma},t)}
                + 2N
                \dyad{w_{v,\text{NN}}}
                \braket{\Phi_{\boldsymbol{j}}(\boldsymbol{\gamma},t)}
                \\&\quad
                + \sqrt{2N}
                \dyad{w_{v,\boldsymbol{j_0}}}{w_{v,\text{NN}}}
                \braket{\Phi_{\boldsymbol{j}}(\boldsymbol{\gamma},t)}
                {\Phi_{\boldsymbol{j_0}}(\boldsymbol{\gamma},t)}
                + \sqrt{2N}
                \dyad{w_{v,\text{NN}}}{w_{v,\boldsymbol{j_0}}}
                \braket{\Phi_{\boldsymbol{j_0}}(\boldsymbol{\gamma},t)}
                {\Phi_{\boldsymbol{j}}(\boldsymbol{\gamma},t)},
    \end{aligned}
\end{equation}
where we have that
\begin{align}
    &\braket{\Phi_{\boldsymbol{j_0}}(\boldsymbol{\gamma},t)}
        = \abs{\braket{\boldsymbol{\gamma}}{\boldsymbol{\chi}_{q\neq 1}(t)}}^2,
    \\
    &\braket{\Phi_{\boldsymbol{j}}(\boldsymbol{\gamma},t)}
        = \abs{h_1(t,t_0)}^2 + \abs{h_2(t,t_0)}^2,
    \\
    &\braket{\Phi_{\boldsymbol{j_0}}(\boldsymbol{\gamma},t)}{\Phi_{\boldsymbol{j}}(\boldsymbol{\gamma},t)}
        = \braket{\boldsymbol{\gamma}}{\boldsymbol{\chi}_{q\neq 1}(t)}^*
        h_2(t,t_0).
\end{align}

\subsection{Entanglement characterization after the \emph{conditioning to HHG} operation}
Here, we look at the state obtained after the \emph{conditioning to HHG} operation, and proceed to characteriz the light-matter entanglement in a similar basis as what it was done before. In this case, we have
\begin{equation}
    \begin{aligned}
    \braket{\boldsymbol{\gamma}}{\Psi(t)}
        &= D(\alpha)
            \bigg[
                \ket{w_{v,\boldsymbol{j}_0}}
                \braket{\boldsymbol{\gamma}}{\bar{\Phi}_{\boldsymbol{j_0}}(t)}
                + \sum_{\boldsymbol{j}\in\text{NN}}
                    \ket{w_{v,\boldsymbol{j}}}
                    \braket{\boldsymbol{\gamma}}{\bar{\Phi}_{\boldsymbol{j}}(t)}
            \bigg]
        \\&
        =D(\alpha)
            \bigg[
                \ket{w_{v,\boldsymbol{j}_0}}
                \ket{\bar{\Phi}_{\boldsymbol{j_0}}(\boldsymbol{\gamma},t)}
                + \sum_{\boldsymbol{j}\in\text{NN}}
                    \ket{w_{v,\boldsymbol{j}}}
                    \ket{\bar{\Phi}_{\boldsymbol{j}}(\boldsymbol{\gamma},t)}
            \bigg],
    \end{aligned}
\end{equation}
where we have defined
\begin{equation}
    \ket{\bar{\Phi}_{\boldsymbol{j}}(\boldsymbol{\gamma},t)}
        = \braket{\boldsymbol{\gamma}}{\bar{\Phi}_{\boldsymbol{j}}(t)}.
\end{equation}

Similarly to what we did before, in order to compute the entropy of entanglement we perform the partial trace with respect to the quantum optical degrees of freedom. We thus have
\begin{equation}
    \begin{aligned}
    \bar{\rho}_{\text{elec}}(t)
        &= 
                \dyad{w_{v,\boldsymbol{j_0}}}
                \braket{\bar{\Phi}_{\boldsymbol{j_0}}(\boldsymbol{\gamma},t)}
                + 2N
                \dyad{w_{v,\text{NN}}}
                \braket{\bar{\Phi}_{\boldsymbol{j}}(\boldsymbol{\gamma},t)}
                \\&\quad
                + \sqrt{2N}
                \dyad{w_{v,\boldsymbol{j_0}}}{w_{v,\text{NN}}}
                \braket{\bar{\Phi}_{\boldsymbol{j}}(\boldsymbol{\gamma},t)}
                {\bar{\Phi}_{\boldsymbol{j_0}}(\boldsymbol{\gamma},t)}
                + \sqrt{2N}
                \dyad{w_{v,\text{NN}}}{w_{v,\boldsymbol{j_0}}}
                \braket{\bar{\Phi}_{\boldsymbol{j_0}}(\boldsymbol{\gamma},t)}
                {\bar{\Phi}_{\boldsymbol{j}}(\boldsymbol{\gamma},t)},
    \end{aligned}
\end{equation}
where we have that
\begin{align}
    &\braket{\bar{\Phi}_{\boldsymbol{j_0}}(\boldsymbol{\gamma},t)}
        = 1-\abs{\braket{\bar{0}}{\chi(t,t_0)}}^2,
    \\
    &\braket{\bar{\Phi}_{\boldsymbol{j}}(\boldsymbol{\gamma},t)}
        = -\bar{h}_1(t,t_0)\braket{\bar{0}}{\boldsymbol{\chi}(t,t_0)}^*\braket{\boldsymbol{\gamma}}{\bar{0}_{q\neq 1}}^* \chi_1(t,t_0)^* e^{-\frac12\abs{\chi_1(t,t_0)}^2}
        + \bar{h}_2(t,t_0) \braket{\boldsymbol{\gamma}}{\boldsymbol{\chi}_{q\neq 1}(t,t_0)}\nonumber
        \\&\hspace{3.2cm}
        - \bar{h}_2(t,t_0)
        \braket{\bar{0}}{\boldsymbol{\chi}(t,t_0)}^*
        \braket{\boldsymbol{\gamma}}{\bar{0}_{q\neq 1}}^*e^{-\frac12\abs{\chi_1(t,t_0)}^2}
        + \bar{h}_3(t,t_0)\braket{\boldsymbol{\gamma}}{\boldsymbol{\chi}_{q\neq 1}(t,t_0)}
        e^{-\frac12\abs{\chi_1(t,t_0)}^2}\nonumber
        \\&\hspace{3.2cm}
        - \bar{h}_3(t,t_0)
        \braket{\bar{0}}{\boldsymbol{\chi}(t,t_0)}^*
        \braket{\boldsymbol{\gamma}}{\bar{0}_{q\neq 1}}^*,
    \\
    &\braket{\bar{\Phi}_{\boldsymbol{j_0}}(\boldsymbol{\gamma},t)}{\bar{\Phi}_{\boldsymbol{j}}(\boldsymbol{\gamma},t)}
        = \abs{h_1(t,t_0)}^2 
            + \abs{h_2(t,t_0)}^2
            + \abs{h_3(t,t_0)}^2\nonumber
            \\&\hspace{3.2cm}
            + \big[
                h_1(t,t_0)^*h_3(t,t_0)
                \chi_1^*(t,t_0)
                +
                h_1(t,t_0)h_3(t,t_0)^*
                \chi_1(t,t_0)
            \big]e^{-\frac12\abs{\chi_1(t,t_0)}^2}\nonumber
            \\&\hspace{3.2cm}
            + \big[
                h_2(t,t_0)^*h_3(t,t_0)
                +
                h_2(t,t_0)h_3(t,t_0)^*
            \big]e^{-\frac12\abs{\chi_1(t,t_0)}^2}.
\end{align}
\bibliography{References.bib}{}